\documentclass[namedreferences,hyperref,optionalrh]{spr-sola}
\usepackage{graphicx}        % For eps figures, newer & more powerfull
\usepackage{color}           % For color text: \color command
\chardef\us=`\_

\newcommand{\Lyns}{Ly$\alpha$}
\newcommand{\Ly}{Ly$\alpha\;$}

\newcommand{\comment}{}
\newcommand{\commentend}{}
\usepackage{tabularx} %FOR THEAD COMMAND
\usepackage{makecell} 
\usepackage[T1]{fontenc} %FIXES ERROR IN BIBLIOGRAPHY
\usepackage{anyfontsize} %REMOVES THREE YELLOW ERRORS
\usepackage{booktabs} %ADDITIONAL LINES FOR TABLES
\usepackage{textcomp,gensymb}

\mathchardef\mhyphen="2D

\graphicspath{{Figures_Mockup}}

\usepackage{geometry}
\begin{document}

\begin{article}
\begin{opening}
\title{Eruption-Related Ultraviolet Irradiance Enhancements Associated with Flares}

\author[addressref=aff1,corref,email={lmajury01@qub.ac.uk}]{\inits{L.H.}\fnm{Luke}~\lnm{Majury}\orcid{0009-0002-8491-9593}}%\sep

\author[addressref=aff2,corref,email={marie.dominique@oma.be}]{\inits{M.}\fnm{Marie}~\lnm{Dominique}\orcid{0000-0002-1196-4046}}%\sep

\author[addressref=aff1,corref,email={r.milligan@qub.ac.uk}]{\inits{R.O.}\fnm{Ryan}~\lnm{Milligan}\orcid{0000-0001-5031-1892}}%\sep

\author[addressref=aff2,corref,email={dana.talpeanu@oma.be}]{\inits{D.-C.}\fnm{Dana-Camelia}~\lnm{Talpeanu}\orcid{0000-0002-9311-9021}}%\sep

\author[addressref=aff2,corref,email={ingolf.dammasch@oma.be}]{\inits{I.E.}\fnm{Ingolf}~\lnm{Dammasch}}%\sep

\author[addressref=aff2,corref,email={david.berghmans@oma.be}]{\inits{D.}\fnm{David}~\lnm{Berghmans}\orcid{0000-0003-4052-9462}}%\sep

\address[id=aff1]{Astrophysics Research Centre, School of Mathematics and Physics, Queen's University Belfast, University Road, BT7 1NN, Northern Ireland, UK}

\address[id=aff2]{SIDC, Royal Observatory of Belgium, 3 Avenue Circulaire, 1180 Uccle, Belgium}

\runningauthor{Majury et al.}
\runningtitle{Eruption-Related Ultraviolet Irradiance Enhancements Associated with Flares}

\begin{abstract}
    Large solar flares (GOES M-class or higher) are usually associated with eruptions of material. However, when considering flare irradiance enhancements and dynamics such as chromospheric evaporation, potential contributions from erupted material have historically been neglected. We analyse nine eruptive M- and X-class flares from 2024 to early 2025, quantifying the relative contributions of erupted material to irradiance enhancements during the events. SDO/AIA images from four different channels had ribbon and eruption irradiance contributions separated using a semi-automated masking method. The sample-averaged percentages of excess radiated energy by erupted material over the impulsive phase were $10^{+4}_{-4}\%$, $24^{+14}_{-14}\%$, $21^{+14}_{-10}\%$ and $13^{+6}_{-9}\%$ for the $131\,\textrm{Å}$, $171\,\textrm{Å}$, $304\,\textrm{Å}$ and $1600\,\textrm{Å}$ channels, respectively. For three events that were studied in further detail, HXR imaging showed little to no signatures of nonthermal heating within the eruptions. Our results suggest that erupted material can be a significant contributor to UV irradiance enhancements during flares, with possible heating mechanisms including nonthermal particle heating, Ohmic heating, or dissipation of MHD waves. Future work may clarify the heating mechanism and evaluate the impact of eruptions on spectral variability, particularly in Sun-as-a-star and stellar flare observations.
\end{abstract}

\end{opening}
%-------------------------------------------------
    \section{Introduction}

    \comment Flares are transient bursts of electromagnetic radiation from the solar atmosphere, these events are often associated with eruptions; dynamic ejections of material from either prominences suspended in the corona or directly from the lower layers of the solar atmosphere \citep{Webb_2012,Raouafi_2016,Shen_2021}. Both flares and eruptions occur due to the destabilisation of magnetic flux in the solar corona, often following the emergence of flux from the solar interior\commentend. In many cases, instabilities in this flux lead to the eruption of a magnetic flux rope, in the process of which a current sheet is formed \citep[e.g.][]{Heyvaerts_1977,Antiochos_1999,Moore_2001}. As the current sheet thins, the assumption of ideal magnetohydrodynamics (MHD) break down (specifically the frozen-in flux condition), resulting in fast magnetic reconnection. This releases magnetic free energy, exceeding $10^{32}\,\textrm{erg}$ in the largest events. The released energy is partitioned into plasma heating, bulk motions of eruptions, and acceleration of particles to nonthermal energies \citep{Fletcher_2011,Emslie_2012,Aschwanden_2014}. Though not all flares have associated coronal mass ejections (CMEs)\comment, eruptions that escape the Sun's low coronal magnetic field\commentend, the majority of large flares (GOES M-class or larger) are associated with these successful eruptions, largely due to the magnetic structure of their active regions \citep{Yashiro_2006,Wang_2007,Inoue_2013,Zhang_2022,Li_2024}.
    
    The radiated energy during major flares can, though infrequently, exceed $10^{32}\,\textrm{erg}$ in ultraviolet (UV) emission alone \citep{Woods_2006}. This emission originates primarily from chromospheric flare ribbons and coronal loops, and constitutes a large fraction of the total radiated energy \citep{Milligan_2014,Kontar_2017,Warmuth_2020}. Such UV emissions are known to influence the Earth's ionosphere, ionising neutral particles and driving currents that generate magnetic perturbations, which are observed as a Solar Flare Effect \citep[Sfe;][]{Mitra_1974,Curto_2020}. Additionally, the increased ionisation of the ionosphere can interfere with radio communications, as variations in plasma frequency alter the height of the Earth-ionosphere waveguide \citep{Cannon_2013}. While the majority of excess UV emissions during flares originate from ribbons and loops, a significant amount of radiation may also originate from eruptions associated with flares. For example, \citet{Rubiodacosta_2009} studied an M1.4 class flare with an associated eruption in \Ly images, finding the radiated power from the flare footpoints to be on the order of $10^{26}\,\textrm{erg\,s}^{-1}$, corresponding to less than $10\%$ of the available nonthermal power in accelerated electrons. While the associated eruption had a much lower surface intensity of $1.2\times10^{6}\,\textrm{erg\,cm}^{-2}\textrm{\,s}^{-1}$ than the footpoints at $6.7\times10^{7}\,\textrm{erg\,cm}^{-2}\textrm{\,s}^{-1}$, the area of erupted material was much larger, suggesting that it may have contributed significantly to the total flare excess in \Lyns. Contributions of erupted material to the overall \Ly flare excess have also been suggested for flares studied by \citet{Milligan_2021} and \citet{Wauters_2022}. \citet{Mierla_2022} reported a bright prominence eruption in He~\textsc{ii} $304\,\textrm{Å}$ emission, with the authors attributing its brightness to collisional excitation processes within the prominence. Further analysis by \citet{Hayes_2024} of the same event using X-ray observations from the Spectrometer Telescope for Imaging X-rays on Solar Orbiter \citep[SolO/STIX;][]{Krucker_2020} revealed the prominence eruption to be broadly cospatial with soft X-rays (SXRs) of $4-10\,\textrm{keV}$, with a smaller region of hard X-ray (HXR) emission being seen in the prominence at lower altitudes. From this, the authors inferred heating of the plasma by nonthermal electrons travelling upward from the site of their acceleration. Similar observations of HXR sources associated with eruptions have been reported by \citet{Kane_1992}, \citet{Hudson_2001}, \cite{Krucker_2007}, \citet{Glesener_2013}, and \citet{Lastufka_2019}. Understanding the heating mechanisms (e.g. nonthermal particle heating) driving eruption emission may allow the results of models to be better constrained, with insights into the partitioning of magnetic energy release into erupted and chromospheric plasma being provided \citep{Janvier_2015,Dahlin_2025}.
    
    Contributions of erupting material to the overall flare excess have also been discussed in the context of spectral observations, being suggested as a potential cause of observed blue asymmetries and Doppler shifts, as opposed to chromospheric evaporation \citep[e.g.][]{Batchelor_1991,Ding_2003,Majury_2025,Majury_2025_1}. Observations of spectral variability in spatially integrated observations of stellar flares have previously been attributed to both chromospheric evaporation and eruptions \citep{Gunn_1994,Berdyugina_1999,Wang_2024,Argiroffi_2019}. However, recent work by \citet{Dewilde_2025} synthesised Sun-as-a-star spectra from imaging spectroscopy observations using the Numerical Empirical Sun-as-a-Star Integrator code \citep[NESSI;][]{Pietrow_2024}, finding that spectral variations due to gravitationally-bound upflows could be misinterpreted as signatures of Coronal Mass Ejections (CMEs) in spatially integrated data. Understanding the irradiance contributions of eruptions during flares may help guide future interpretations of spectral variability in spatially integrated flare observations.

    In this work, we aim to determine the typical contribution of flare-associated eruptions to the total excess radiated energy. Section \ref{section2} provides an overview of the flare identification process, the instruments used, and the methods applied to separate the contributions of flare and eruption emissions to total irradiance enhancement. Section \ref{section3} provides results of this analysis, including case studies of specific events and statistical information for a larger sample. Section \ref{section4} provides a discussion of, and conclusions from, these results and the broader implications of this work for future research.
        
    \section{Observations and Analysis}
    \label{section2}
        \subsection{Flare Identification}
        
        Flares were initially chosen for study based on the availability of observations from the \Ly High Resolution Imager telescope of the Extreme Ultraviolet Imager on Solar Orbiter \citep[SolO/EUI HRI$_{\mathrm{Lya}}$;][]{Muller_2020,Rochus_2020}. This instrument was selected due to previous observations of bright eruptions being made in \Ly \citep{Rubiodacosta_2009,Wauters_2022}. This resulted in the identification of two flares, an M2.0 event on 2 March 2022 and an M2.3 event on 7 December 2024. Of these two events, only the M2.0 was eruptive. Thus, to provide more robust statistics, eruptive events that were observed by the \Ly channel of the Large Yield Radiometer on the Project for Onboard Autonomy 2 \citep[PROBA-2/LYRA;][]{Hochedez_2006,Dominique_2013} were sought out. A list of M- and X-class events captured by the instrument's backup unit in campaigns during 2024 and early 2025 was compiled. Each event was compared to a list of eruptions identified by the Eruption Patrol software module \citep{Hurlburt_2015} from the Heliophysics Event Knowledgebase \citep[HEK;][]{Hurlburt_2012}. A subset of flares that had an eruption that was between $50^{\prime\prime}$ and $200^{\prime\prime}$ from the flare's location and occurred during the Geostationary Operational Environmental Satellite (GOES) flare period was then identified. Each of 82 events in this subset was then visually inspected in $304\,\textrm{Å}$ images from the Atmospheric Imaging Assembly on the Solar Dynamics Observatory \citep[SDO/AIA;][]{Pesnell_2012,Lemen_2012} using the JHelioviewer visualisation software package \citep{Muller_2017}. For each event, the following criteria were applied:

        \begin{itemize}
        \item Flare had an associated eruption that appeared brighter than the quiet Sun
        \item Eruption must have originated from the flare's active region
        \item Flare ribbons and eruption must have had minimal spatial overlap 
        \end{itemize}

        \noindent
        Applying these criteria resulted in the generation of a final list of nine events (including the event observed by HRI$_{\mathrm{Lya}}$) for which the radiative properties of the flare ribbons and associated eruption emission were analysed in detail. Details of these events are listed in Table \ref{table6.05}.

        \begin{table}[ht]
        \begin{tabular}{ccccccc}
        \hline
        \textbf{Class} & \textbf{Date} & \textbf{Start Time} & \textbf{Peak Time} & \textbf{End Time} & \textbf{\shortstack{Stonyhurst\\Coordinate ($\degree$)}} & \textbf{\shortstack{NOAA AR\\Number}}\\
        \hline
        M1.1 & 2024-03-23 & 06:47 & 06:55 & 06:59 & S13E04 & 13615 \\
        M1.1 & 2025-01-07 & 22:35 & 23:05 & 23:42 & S20W88 & 13939 \\
        M1.9 & 2024-12-21 & 00:33 & 00:38 & 00:42 & S15E61 & 13932 \\
        M2.0 & 2022-03-02 & 17:31 & 17:39 & 17:47 & N15E29 & 12958 \\
        M2.7 & 2025-01-24 & 20:48 & 21:04 & 21:17 & S06W67 & 13961 \\
        M3.8 & 2024-12-19 & 15:27 & 15:34 & 15:39 & S14E81 & 13928\\
        X1.0 & 2024-05-12 & 16:11 & 16:26 & 16:38 & S18W72 & 13664 \\
        X2.0 & 2024-10-31 & 21:12 & 21:20 & 21:27 & N15E28 & 13878 \\
        X9.0 & 2024-10-03 & 12:08 & 12:18 & 12:27 & S15W03 & 13842 \\
        \hline
        \end{tabular}
        \caption{GOES start, peak and end times along with coordinates and active region numbers for the sample of nine eruptive events}
        \label{table6.05}
        \end{table}
        
        \subsection{Separation of Eruption and Ribbon Contributions in UV Images}
            
            \subsubsection{UV Imagers}
            Data from three UV imaging instruments were included in this analysis. Firstly, and primarily, data from SDO/AIA were used. AIA provides high-cadence ($12\,\textrm{s}$) imaging of the entire solar disk in 10 different wavelength bands, covering a wide range of temperatures and hence probing various regions of the solar atmosphere. These specifications enable AIA to provide near-complete coverage of solar flares at a cadence suitable for studying their dynamics. In this work, we employed AIA observations of nine flares with associated bright eruptions, using a mask to spatially separate ribbon and eruption contributions to total flare excess radiated energy. 

            For a single eruptive event, an M2.0 flare on 2 March 2022, further imaging from the high-resolution \Ly and Extreme Ultraviolet (EUV) telescopes of SolO/EUI was used to supplement AIA data. HRI$_{\mathrm{Lya}}$ and HRI$_{\mathrm{EUV}}$ provide images at a resolution of $\sim3^{\prime\prime}$ and $\sim1^{\prime\prime}$, respectively. The absolute resolution (km) of each instrument varies with the orbital radius of SolO. Furthermore, HRI$_{\mathrm{Lya}}$ experiences temperature-driven degradation to its resolution, particularly near perihelion \citep{Berghmans_2023}. Fortunately, the eruptive event that was observed occurred while SolO was near aphelion.
            
            Additionally, for an occulted M1.1 flare on 7 January 2025, images from the Extreme Ultraviolet Imager of the Sun-Earth Connection Coronal and Heliospheric Investigation on the Solar-Terrestrial Relations Observatories \citep[STEREO/SECCHI EUVI;][]{Kaiser_2008,Howard_2008,Wuesler_2004} were used to analyse emissions from the flare ribbons, which were visible from STEREO-A. The EUVI instrument operates with a modest cadence, as fast as $2.5\,\textrm{minutes}$, in four wavelength channels: $171\,\textrm{Å}$, $195\,\textrm{Å}$, $284\,\textrm{Å}$ and $304\,\textrm{Å}$.

            %\subsubsection{GOES/XRS}
            %The X-ray Sensor \citep[XRS;][]{Chamberlin_2009} instrument on GOES provides measurements of soft X-rays in its short ($0.5-4\,\textrm{Å}$) and long channels ($1-8\,\textrm{Å}$). The flux in the long channel is used to classify flare magnitude and establish the start, peak and end times of flares. The flare start time precedes a continuous increase of flux over four consecutive minute-long intervals. The peak time is when the maximum flux following the start time is observed, with the end time being when the flux declines to half of this maximum value. In this work, we primarily use XRS observations as a metric for flare timing, allowing standardised flare and eruption radiated energy measurements between events.
            
            \subsubsection{Ribbon and Eruption Masking}
\begin{figure}[ht] 
                \centerline{
                       \includegraphics[width=\textwidth,clip=]{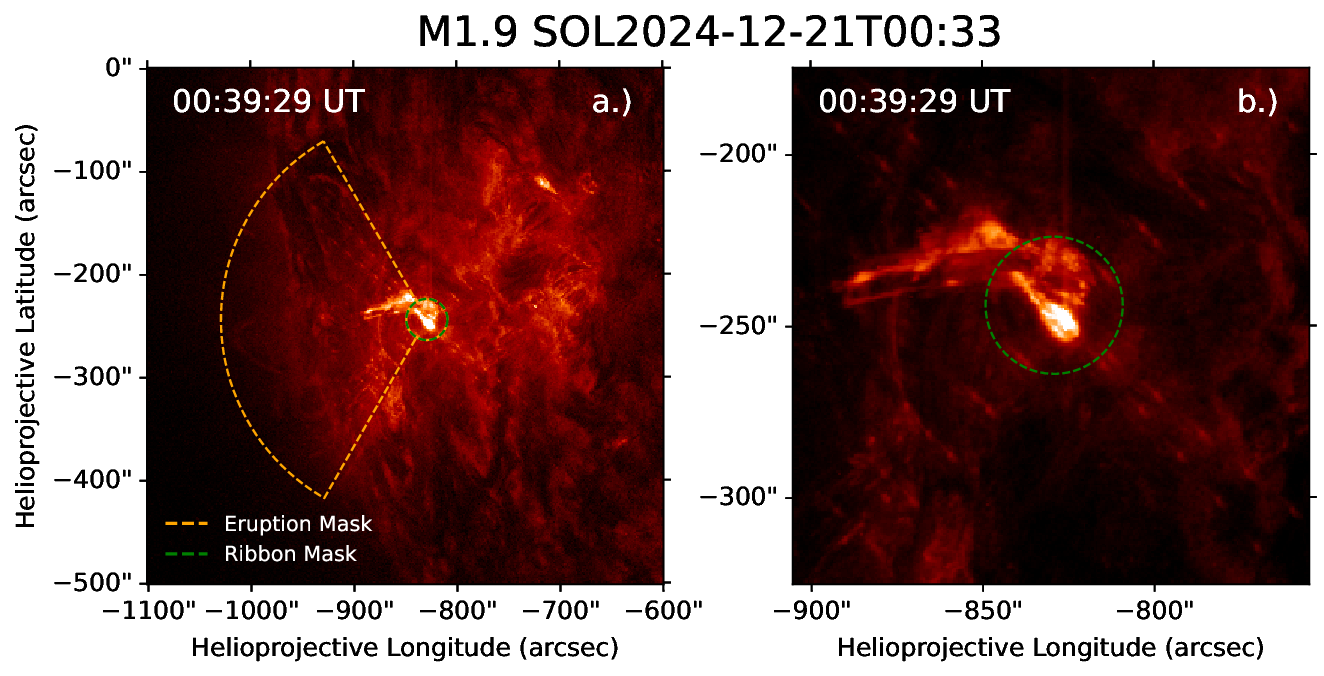}
                      }
            \caption{AIA $304\,\textrm{Å}$ images of M1.9 flare on 21 December 2024 with eruption and ribbon masks overlaid. Panel a.) shows a global view of the event. A close-up of the flare and eruption is shown in panel b.).}
            \label{figure6.12}
            \end{figure}

            For each event in the sample, masks were constructed around the AIA $304\,\textrm{Å}$ frame with the highest count rate ($\textrm{DN\,s}^{-1}$) during the flare. Ribbon masks were defined as a circle sharing a centre with the largest $20\%$ intensity contour, with a radius covering the entirety of the ribbons determined by visual inspection. The eruption masks were defined as a circular sector, sharing a common centre with the corresponding ribbon mask, extending from the edge of the ribbon mask to a manually chosen upper radius. The upper and lower angles of this sector were similarly manually defined. The manual choices in mask selection were made conservatively. Thus, regions with overlapping eruption and ribbon emission were absorbed into the ribbon mask. An example of the eruption and ribbon masks chosen for an M1.9 flare in the sample is shown in Figure \ref{figure6.12}, a.) shows a global view of the event, while b.) shows a close-up of the eruption and ribbons.

            For each flare, AIA images in four channels ($1600\,\textrm{Å}$, $304\,\textrm{Å}$, $171\,\textrm{Å}$ and $131\,\textrm{Å}$) from 30 minutes before the GOES start time to 30 minutes after the GOES end time were analysed. This allowed the relative contribution of eruptions to emissions at different formation temperatures to be ascertained. The ribbon and eruption masks were applied to each image in each channel, with the count rates of pixels within each mask being summed. This provided total count rates of both the ribbons and the eruption for each frame. The count rate values for each channel were then radiometrically calibrated using the AIA response function provided by the \texttt{get\_aia\_response} routine in SolarSoft \citep{Freeland_1998}, providing the radiated energy rates for each image in $\textrm{erg}\,\textrm{s}^{-1}$. Total (ribbon plus eruption) radiated energies during the flares' impulsive phases (GOES start to GOES peak time) were then calculated for each wavelength, with radiated energies over the full flare durations (GOES start to GOES end time) additionally being calculated for the $304\,\textrm{Å}$ and $1600\,\textrm{Å}$ channels.

            For the eruptive flare observed by HRI$_{\mathrm{Lya}}$ and HRI$_{\mathrm{EUV}}$, the same masking technique as described for AIA was applied, with the equivalent mask being reprojected to SolO's point of view during the event. Unfortunately, the calibrated response functions for HRI are not yet available, so calculation of radiated energies in absolute units was not possible. As the event observed by EUVI was occulted from the perspective of SDO, a mask was manually reconstructed using the same approach described for AIA above, rather than reprojecting. As with HRI, radiometric calibration was not performed for EUVI, with only the fraction of excess counts in the ribbon and eruption masks being determined. For both EUI and EUVI, light travel time corrections were applied, synchronising their observations with Earth-based instruments.
            
        \subsection{X-Ray Imaging}
            For three events with available data, HXR observations from STIX or the Hard X-ray Imager on the Advanced Space-based Solar Observatory \citep[ASO-S/HXI;][]{Gan_2019,Zhang_2019} were used to generate HXR images. This allowed any spatial overlap between HXRs, which infer heating by nonthermal particles, and erupted material to be identified. STIX provides spectrally-resolved imaging of solar X-rays with energies between $4$ and $150\,\textrm{keV}$. The instrument employs a Fourier imaging technique, with incident X-ray flux being spatially modulated in a unique Moiré pattern for each of 30 subcollimator-detector pairs. The amplitude and phase of each measured pattern provide Fourier visibilities from which images can be reconstructed. HXI provides similar spectrally-resolved imaging of solar X-rays from Earth orbit. HXI also employs a spatially modulated Fourier imaging technique, with 91 detector-subcollimator pairs. The maximum entropy method \citep[MEM\_GE;][]{Massa_2020} was applied to STIX data to generate images, with the CLEAN algorithm \citep{Hogbom_1974} being applied to HXI data.
        
        \subsection{UV Photometry}  
            In this study, photometric data from PROBA-2/LYRA and the B-channel of the Extreme Ultraviolet Sensor of the Extreme Ultraviolet and X-ray Irradiance Sensors on GOES \citep[GOES/EXIS EUVS-B;][]{Eparvier_2009} were used to determine whether observed irradiance contributions from bright eruptions can be seen in disk-integrated data. This was examined by comparing spatially resolved AIA $304\,\textrm{Å}$ data and spatially integrated observations from EUVS-B and LYRA.
    
            EUVS-B provides moderate-cadence ($30\,\textrm{s}$) observations of solar irradiance between $1180\;\textrm{Å}-1270\;\textrm{Å}$, dominated by \Ly and Si~\textsc{iii} emission. Its predecessor instruments on the GOES 13, 14, and 15 satellites have been widely employed to study \Ly flare variability \citep{Milligan_2020,Lu_2021,Milligan_2021,Greatorex_2023}. PROBA-2/LYRA provides very high-cadence (nominally $50\,\textrm{ms}$) observations of solar irradiance during flares in four EUV channels: \Lyns, Herzberg, Aluminium and Zirconium. The instrument carries three nearly identical units, with the backup unit (unit 1) of the instrument currently maintaining a relatively high signal in its \Ly channel. Past flare studies using the instrument are presented in \citet{Kretzschmar_2013} and \citet{Wauters_2022}. 
            
            \begin{figure}[ht] 
                \centerline{
                       \hspace*{-0.03\textwidth}
                       \includegraphics[width=.55\paperwidth,clip=]{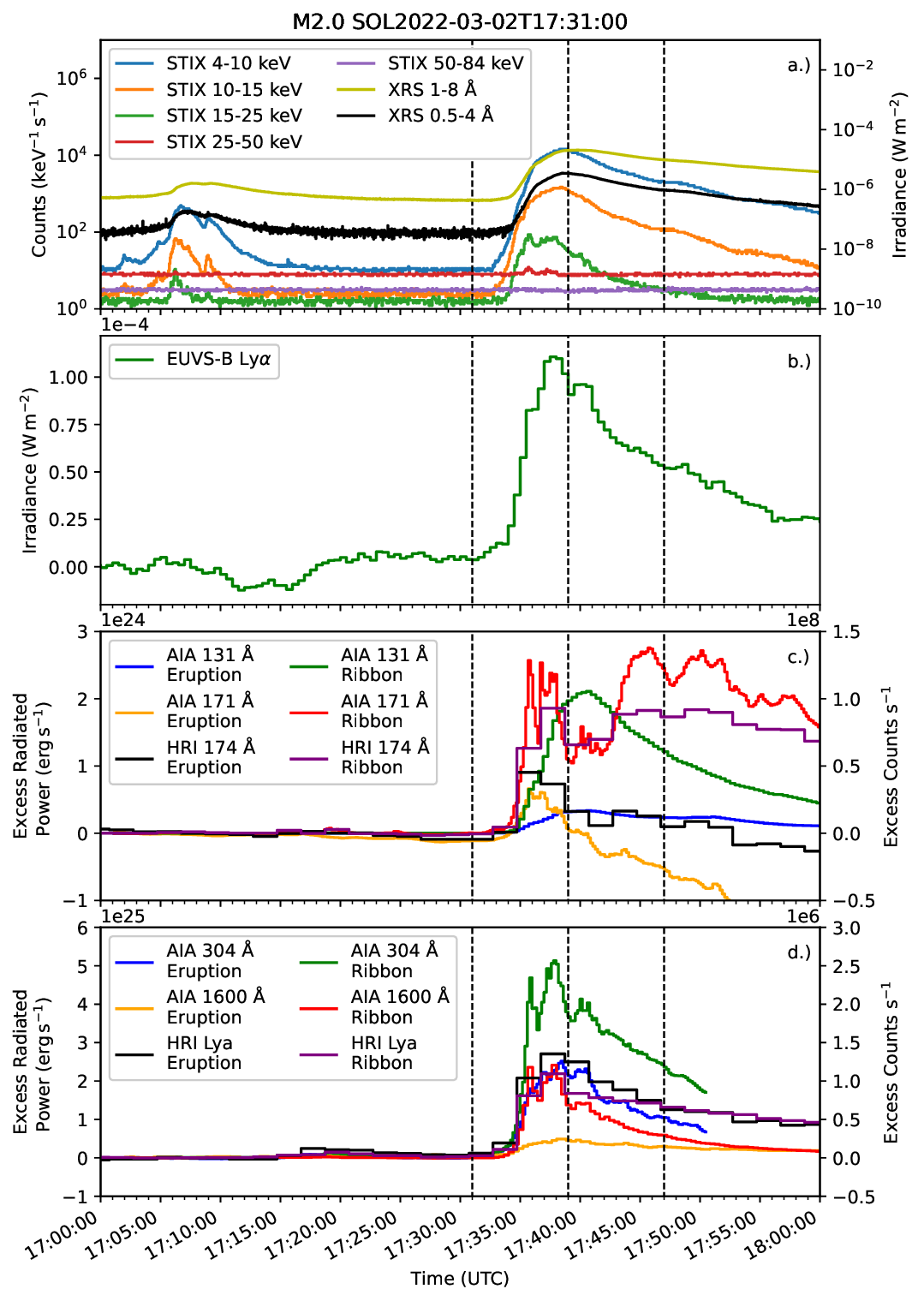}
                      }
            \caption{X-ray and UV emission during M2.0 flare on 2 March 2022. Panel a.) shows lightcurves from STIX and XRS. Excess \Ly emission measured by EUVS-B is displayed in panel b.). Panels c.) and d.) show spatially-separated flare excesses from the flare's eruption and ribbons from four AIA channels (radiated power) and the two channels of HRI (count rate). Black dashed lines indicate GOES start, peak and end times from left to right.}
            \label{figure6.1}
            \end{figure}

            \begin{figure}[ht] 
                \centerline{%\hspace*{0.015\paperwidth}
                       \includegraphics[width=.75\paperwidth,clip=]{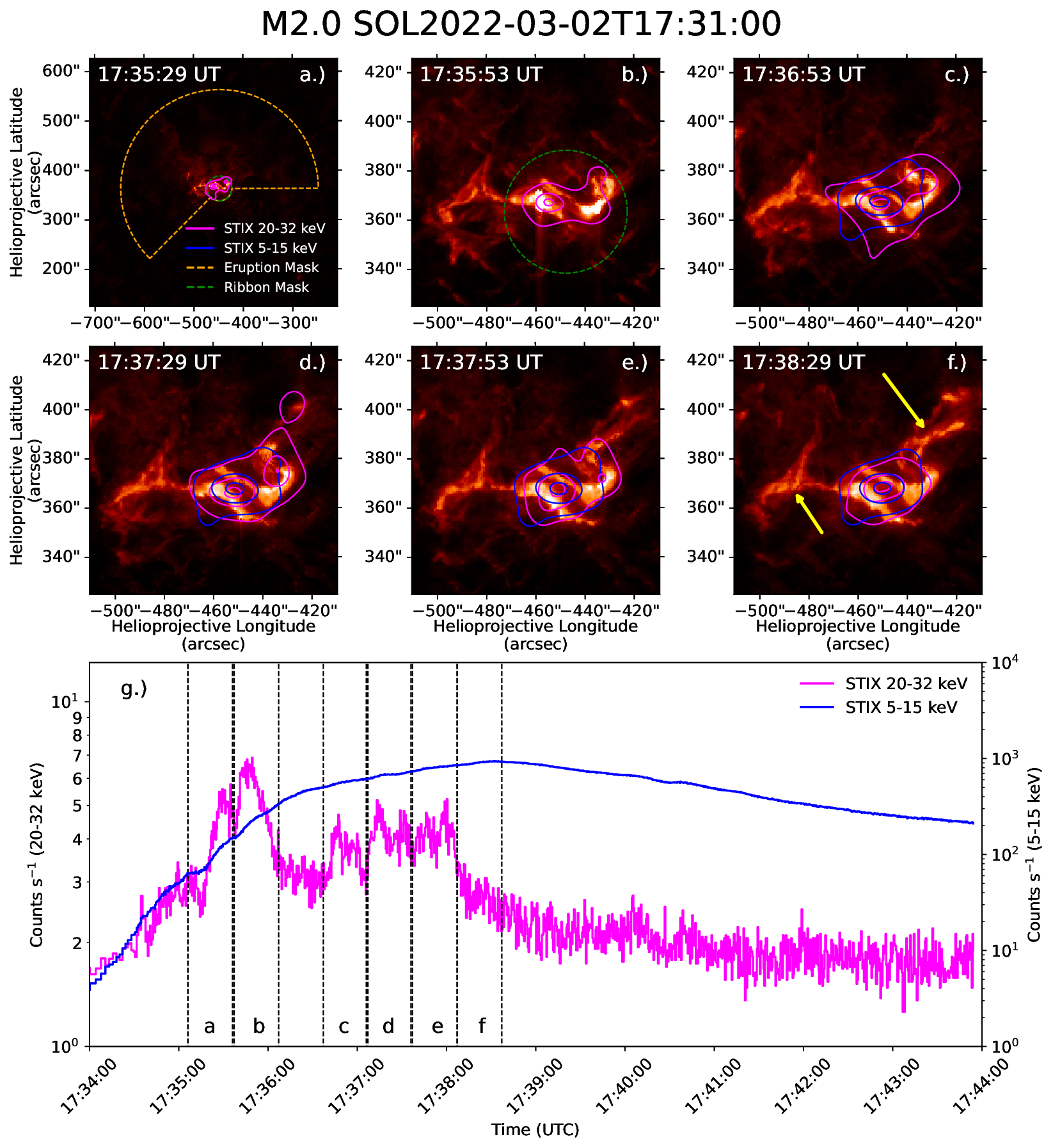}
                      }
            \caption{AIA $304\,\textrm{Å}$ images of M2.0 flare on 2 March 2022 with HXR image contours at 30\%, 50\%, 70\% and 90\% overlaid, shown in panels a.) through f.). Also shown in panel a.) is the mask used to spatially partition the eruption and ribbon contributions to overall flare irradiance enhancement. Panel b.) shows a closer view of the ribbon mask. In panel f.), yellow arrows indicate the locations of erupted material. Panel g.) shows X-ray lightcurves from STIX for emission from $20-32\,\textrm{keV}$ (pink) and $5-15\,\textrm{keV}$ (blue), with the time intervals over which each HXR image in panels a.)-f.) was generated indicated by vertical dashed lines.} 
            \label{figure6.3}
            \end{figure}
        
    \section{Results}
    \label{section3}
        \subsection{The M2.0 Flare on 2 March 2022}
        \label{sectionm2}
    
        % PARAGRAPH THIS OUT BETTER
        HRI$_{\mathrm{Lya}}$ has observed a small number of large flares to date, including an M2.0 flare on 2 March 2022. The event was observed from a similar perspective to Earth-based instruments, with SolO close to aphelion. Lightcurves of the event are presented in Figure \ref{figure6.1}, with X-ray observations from STIX and the X-ray Sensor \citep[XRS;][]{Chamberlin_2009} on GOES presented in panel a.), and \Ly observations from EUVS-B plotted in panel b.). Panel c.) shows excess radiated power from the eruption and ribbons for the $131\,\textrm{Å}$ and $171\,\textrm{Å}$ channels of AIA, as well as the excess count rates for HRI$_{\mathrm{EUV}}$. Similar values for AIA $304\,\textrm{Å}$, and $1600\,\textrm{Å}$, along with HRI$_{\mathrm{Lya}}$ are presented in panel d.). From GOES start to peak time (i.e. the flare's impulsive phase), the percentages of the total flare excess energy radiated by the eruption were $15\%$, $41\%$, $34\%$, $19\%$ for the $131\,\textrm{Å}$, $171\,\textrm{Å}$, $304\,\textrm{Å}$ and $1600\,\textrm{Å}$ channels of AIA, respectively. The fractions of total excess counts in the eruption measured by HRI$_{\mathrm{Lya}}$ and HRI$_{\mathrm{EUV}}$ were $18\%$ and $55\%$, respectively. The timing of peak eruption emission was within $36\,\textrm{s}$ of the ribbons' peak for AIA $304\,\textrm{Å}$. This close temporal overlap resulted in the ribbon and eruption peaks being indistinguishable in photometric \Ly observations, making inference of an eruption impossible from these observations alone.

        \begin{figure}[ht] 
            \centerline{
                   \includegraphics[width=.55\paperwidth,clip=]{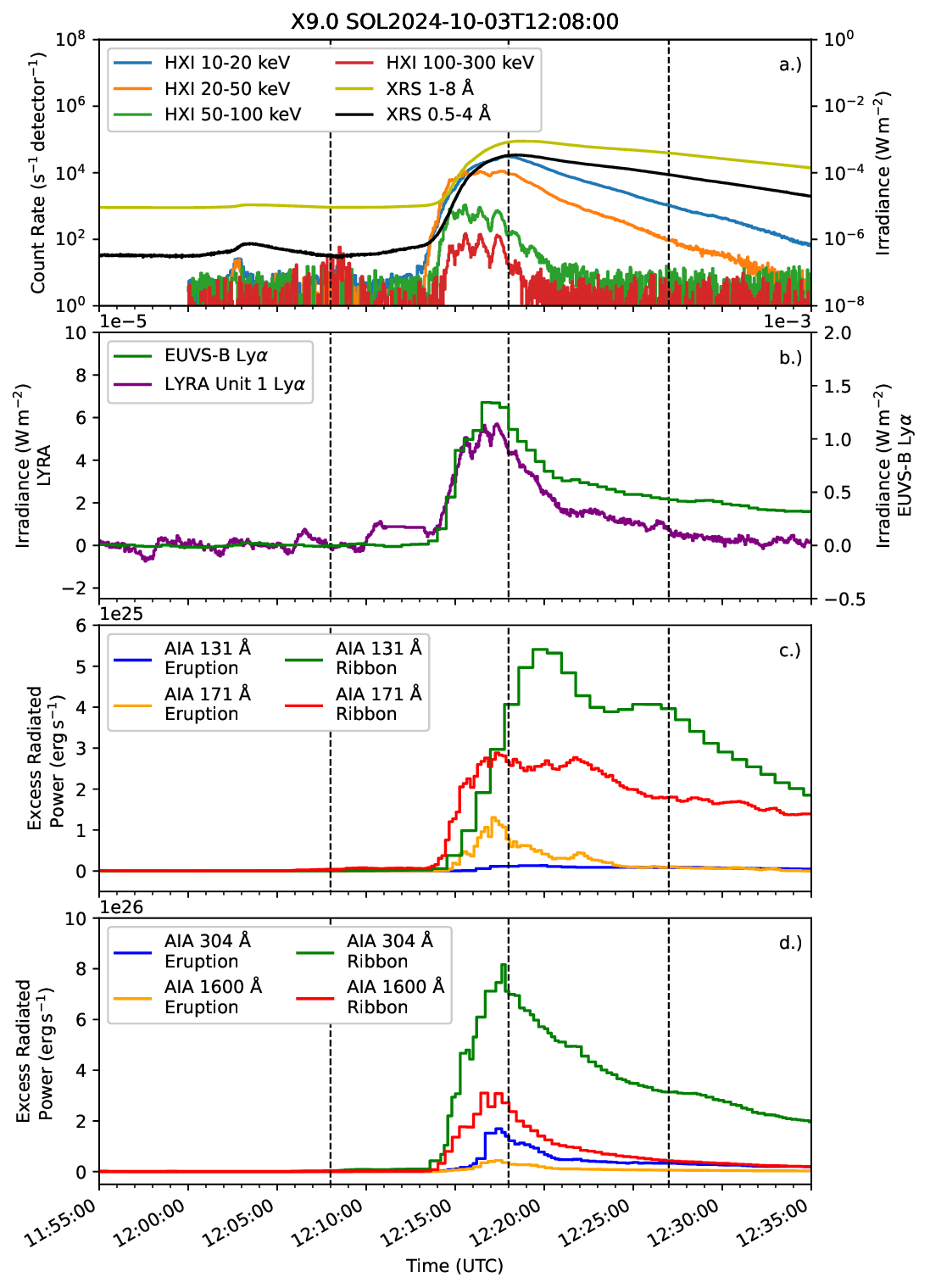}
                  }
        \caption{X-ray and UV emission during X9.0 flare on 3 October 2024. Panel a.) shows lightcurves from HXI and XRS. Excess \Ly emission measured by EUVS-B and LYRA is displayed in panel b.). Panels c.) and d.) show spatially-separated excess radiated power from the flare's eruption and ribbons from four AIA channels. Black dashed lines indicate GOES start, peak and end times from left to right.}
        \label{figure6.6}
        \end{figure}
        
        Figure \ref{figure6.3} shows the evolution of the M2.0 flare and associated eruption in AIA $304\,\textrm{Å}$ images, with MEM\_GE image contours for SXR ($5-15\,\textrm{keV}$) and HXR ($20-32\,\textrm{keV}$) emission overlaid. Each panel represents a different time interval during the event, with panels a.) and b.) showing X-ray emission spatially coincident with ribbons seen in AIA $304\,\textrm{Å}$. In panels c.) and d.), material can be seen erupting near the northmost ribbon, with a HXR source extending in the direction of this eruption. Panel d.) shows a distinct HXR source overlapping with this eruption, indicating nonthermal particle heating within the eruption. However, panels e.) and f.) show the ribbons at peak brightness, with a lack of HXR or SXR emission sources in the bright erupted material. Notably, SXR emission did not overlap with the eruption, indicating its temperature to be relatively cooler than the flare loops. Panel g.) shows nonthermal ($20-32\,\textrm{keV}$) and thermal ($5-15\,\textrm{keV}$) X-ray lightcurves from STIX, with the integration times used to reconstruct each set of images indicated by dashed black lines.
        
        \begin{figure}[ht] 
            \centerline{
                   \includegraphics[width=\textwidth,clip=]{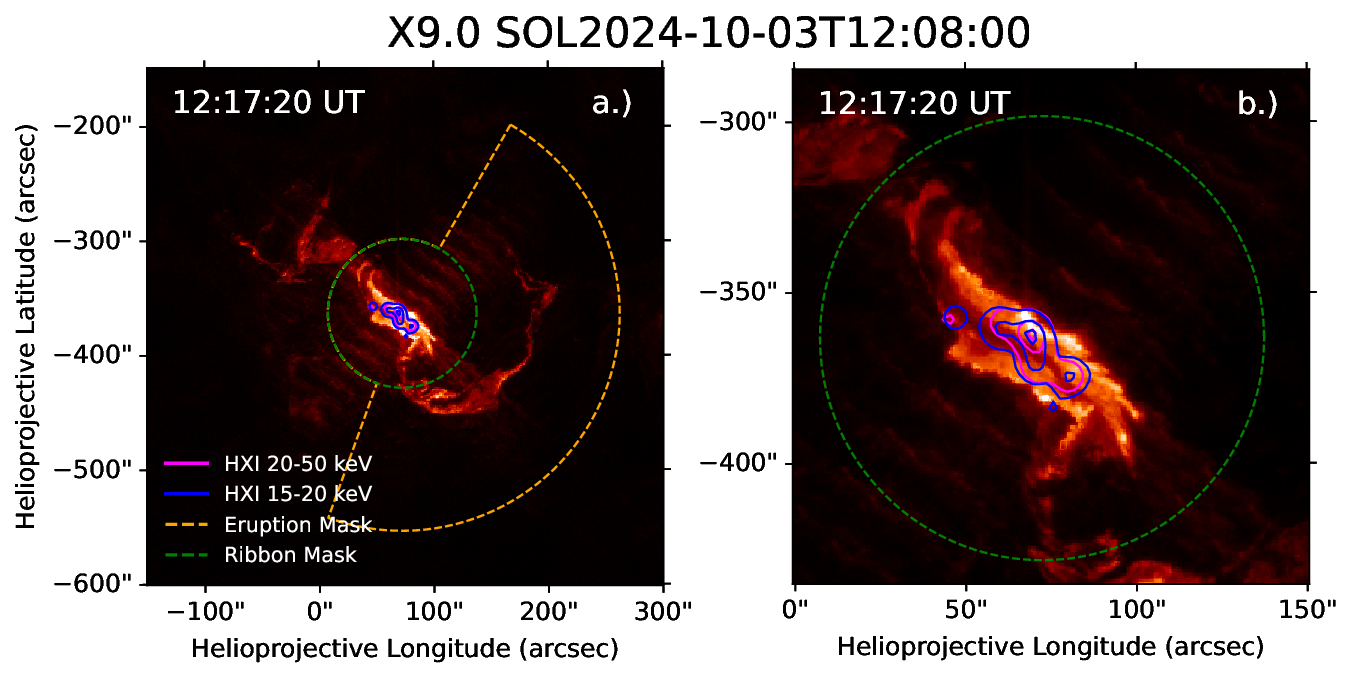}
                  }
        \caption{AIA $304\,\textrm{Å}$ images of X9.0 flare on 3 October 2024 with HXR image contours at 10\%, 50\% and 90\%, and the mask used to spatially partition the eruption and ribbon contributions to overall flare irradiance enhancement overlaid. HXR images were generated for energies of $20-50\,\textrm{keV}$ and $15-20\,\textrm{keV}$ over a time interval of $30\,\textrm{s}$ (12:17:00-12:17:30 UT). Panel a.) shows a global view of the event. A close-up of the flare is shown in panel b.).}
        \label{figure6.8}
        \end{figure}
        
        \subsection{The X9.0 Flare on 3 October 2024}
        To date, the largest flare of Solar Cycle 25 was the X9.0 event on 3 October 2024. Lightcurves for the event in X-ray and UV wavelengths are presented in Figure \ref{figure6.6}, with panel a.) showing X-ray observations from XRS and HXI, and panel b.) displaying \Ly observations from LYRA (unit 1) and EUVS-B. Panel c.) displays eruption and ribbon excess radiated power for AIA $131\,\textrm{Å}$ and $171\,\textrm{Å}$. The equivalent radiated powers for AIA $304\,\textrm{Å}$ and $1600\,\textrm{Å}$ are shown in panel d.). During the impulsive phase, the fractions of total excess energy radiated by the eruption were $7\%$, $22\%$, $12\%$ and $10\%$ for the $131\,\textrm{Å}$, $171\,\textrm{Å}$, $304\,\textrm{Å}$ and $1600\,\textrm{Å}$ channels, respectively. The relatively large fraction of energy radiated in the eruption for the $171\,\textrm{Å}$ channel is likely driven in part by pixel bleed of ribbon emission into the eruption mask, which is mitigated for the $304\,\textrm{Å}$ and $131\,\textrm{Å}$ channels through short exposures. \comment For this reason the event is excluded from statistics of $171\,\textrm{Å}$ emission in Section \ref{section:stats}\commentend. The timing of peak enhancement, as in the M2.0 event, is similar for ribbons and eruption, with the eruption peaking $15\textrm{s}$ before the ribbons in $304\,\textrm{Å}$ emission. This closeness in time of their emission again makes the signatures of eruption emission unclear in \Ly photometric observations. The fast cadence of LYRA (integrated here to $3\,\textrm{s}$) shows three distinct peaks in the impulsive phase, corresponding to three bursts of HXR emission seen in HXI $100-300\,\textrm{keV}$ emission, indicating a likely nonthermal origin to this \Ly flare emission. These bursts illustrate short-period quasi-periodic pulsations in the event, which have been studied in further detail using HXI data in \citet{Li_2025}. Though a halo CME occurs during the event, as identified in the CACTUS archive, little to no coronal dimming is seen in the ribbon or eruption lightcurves in $171\,\textrm{Å}$ emission. 
        
        AIA $304\,\textrm{Å}$ images of the flare are displayed in Figure \ref{figure6.8}, with CLEAN images generated using HXI observations of X-rays between $15-20\,\textrm{keV}$ and $20-50\,\textrm{keV}$ overlaid, and the respective ribbon and eruption masks overplotted. The bright eruption was spatially distinct from the primary HXR sources, suggesting it was likely not strongly heated by nonthermal particles.
        
        \subsection{The M1.1 Flare on 7 January 2025}
        \begin{figure}[ht] 
            \centerline{\hspace*{0.015\paperwidth}
                   \includegraphics[width=.55\paperwidth,clip=]{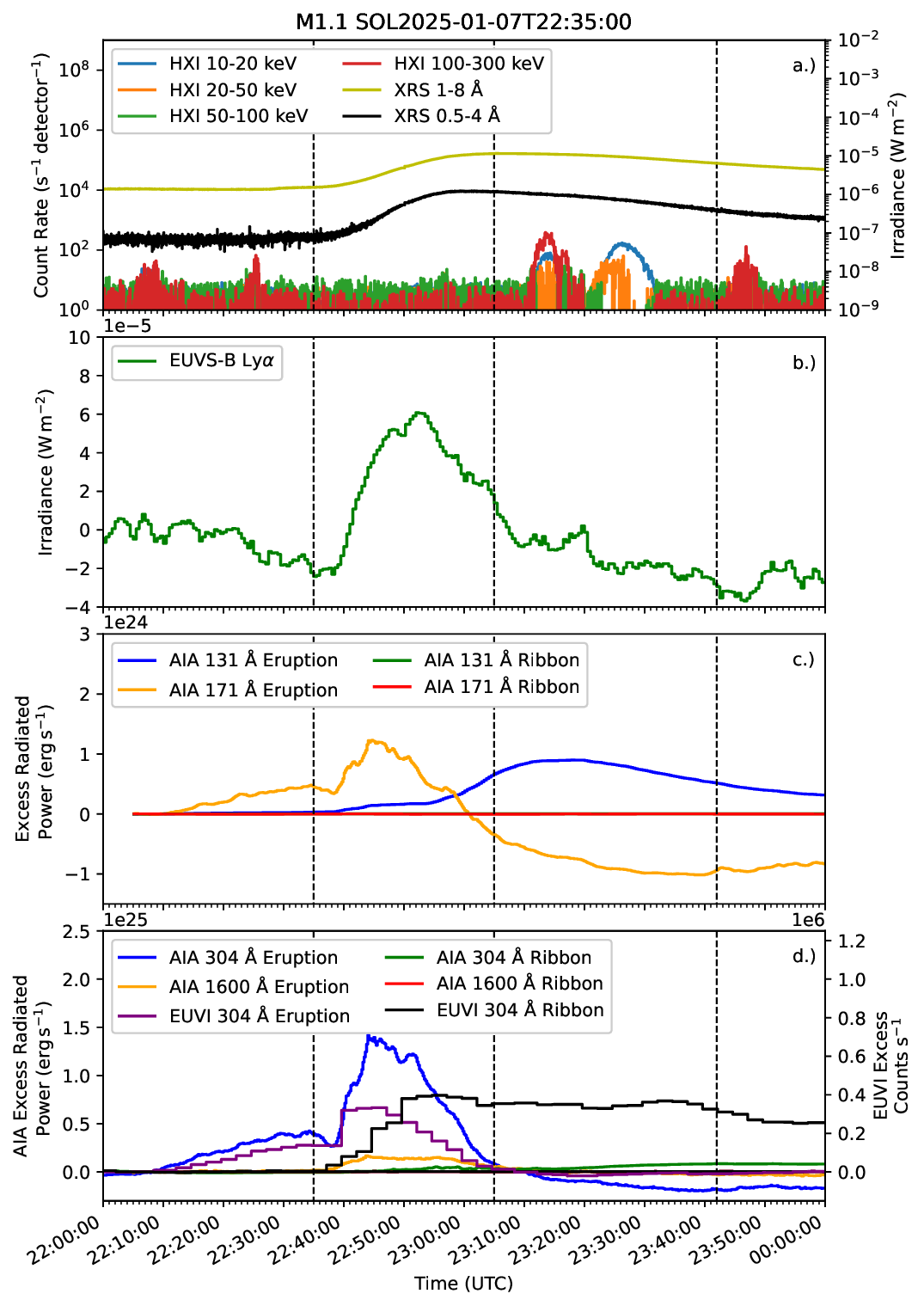}
                  }
        \caption{X-ray and UV emission during M1.1 flare on 7 January 2025. Panel a.) shows X-ray lightcurves from HXI and XRS. Excess \Ly emission measured by EUVS-B is displayed in panel b.). Panels c.) and d.) show spatially-separated excess emission from the flare's eruption and ribbons from four AIA channels (radiated power) and the $304\,\textrm{Å}$ channel of EUVI (count rate). Black dashed lines indicate GOES start, peak and end times from left to right.}
        \label{figure6.9}
        \end{figure}
        The M1.1 flare on 7 January 2025 is the sole behind-the-limb flare of our sample. Figure~\ref{figure6.9} shows lightcurves of the event, with X-ray observations from XRS and HXI shown in panel a.). \Ly observations from EUVS-B are shown in panel b.). Panel c.) shows excess radiated power by the eruption and a quiet-Sun region in AIA $131\,\textrm{Å}$ and $171\,\textrm{Å}$ emission. Similarly, panel d.) shows excess radiated power in AIA $304\,\textrm{Å}$ and $1600\,\textrm{Å}$ emission, along with excess count rates in $304\,\textrm{Å}$ emission measured by EUVI. Over the impulsive phase, the percentage of excess energy radiated by the eruption was close to $100\%$ in each AIA channel, due to the occultation of the ribbons. For EUVI, which observed both eruption and ribbons, $41\%$ of the excess counts were radiated by the eruption over the impulsive phase in $304\,\textrm{Å}$ emission. Assuming a similar fraction for AIA implies an excess radiated energy by the ribbons of $\sim2.5\times10^{28}\,\textrm{erg}$. HXI lightcurves show no HXR emission associated with the eruption, suggesting that the eruption's brightening was not driven by local nonthermal heating. Coronal dimming was seen in $171\,\textrm{Å}$ emission, similarly to the M2.0 event in Section \ref{sectionm2}. A CME associated with the event is listed in the CACTUS archive, indicating that the eruption was successful. EUVI observations show peak ribbon emission at 22:45:53 UT, ten minutes after the peak brightness of the eruption. This behaviour in timing somewhat resembles that seen in flare hot X-ray onsets \citep[e.g.][]{Hudson_2021,Battaglia_2023}. However, the eruption's emission appears relatively cooler than these onsets, which have temperatures $>10^7\,\textrm{K}$, appearing brightest in He~\textsc{ii} $304\,\textrm{Å}$ emission that forms at $\textrm{T}=10^{4.9}\,\textrm{K}$.   

\begin{figure}[ht!] 
            \centerline{
                    \includegraphics[width =0.59 \paperwidth,clip=]{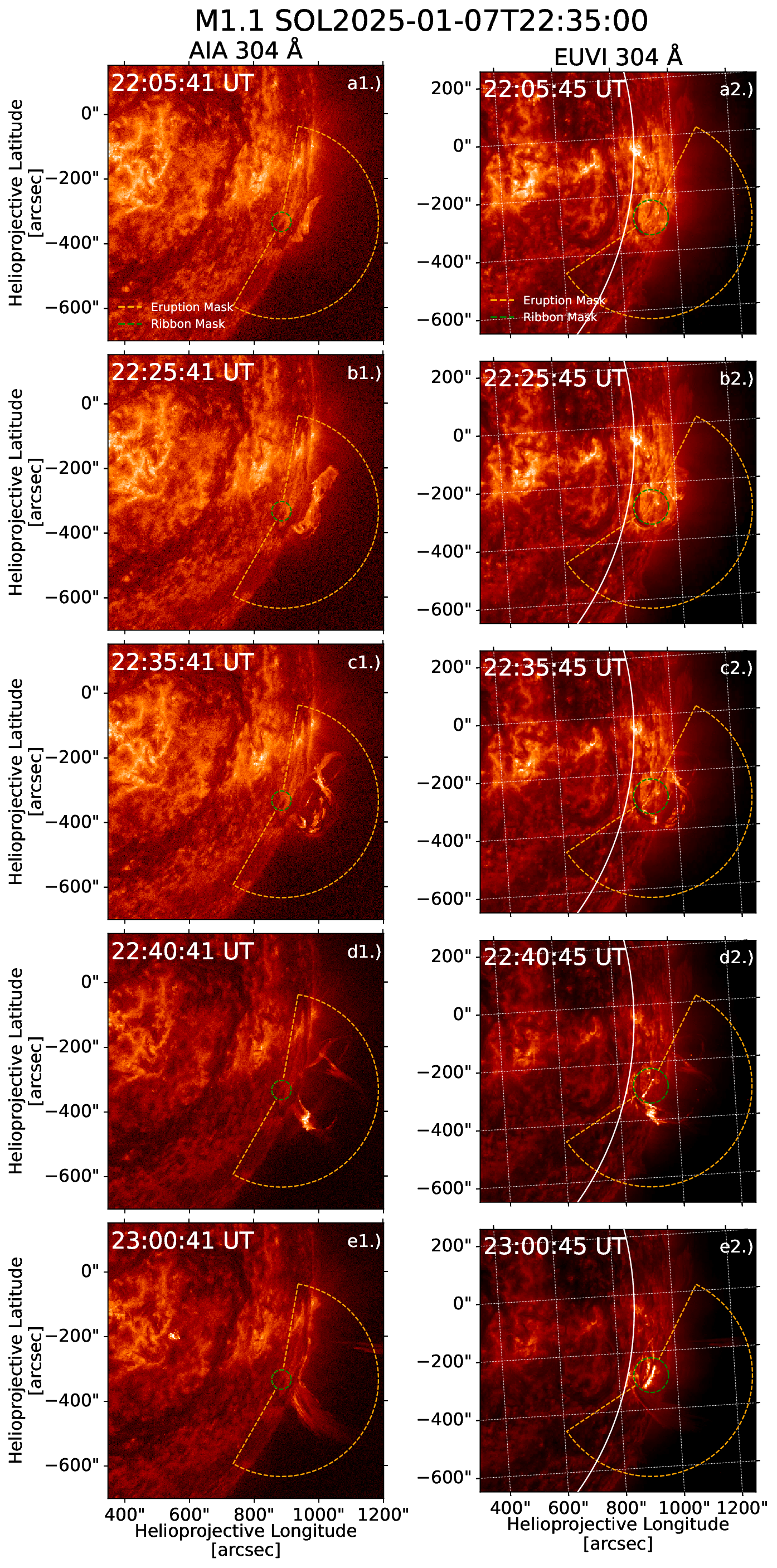}
                  }
        \caption{AIA (left column) and EUVI (right column) $304\,\textrm{Å}$ images of M1.1 flare on 7 January 2025, with the eruption and ribbon masks applied for each set of images overplotted in each panel. The solar limb as seen from AIA is plotted as a white line on each EUVI image.}
        \label{figure6.10}
        \end{figure}

        Images of the flare in $304\,\textrm{Å}$ emission from AIA and EUVI are plotted in the left and right columns of Figure \ref{figure6.10}, respectively, with eruption and ribbon masks for each instrument overplotted. Panels a1.)-b1.) and a2.)-b2.) show an erupting filament, with the structure appearing to unravel in panels c1.) and c2.). The structure shows clear brightening in panels d1.) and d2.), with flare ribbons visible in the EUVI image. e1.) and e2.) show the erupted material post-brightening, with the ribbons appearing strongest in e2.).

        \subsection{Statistics from Nine M- and X-Class Events}
        \label{section:stats}
        \begin{table}[ht]
        \begin{tabular}{cccccc}
        \hline
        \textbf{Class} & \textbf{Start Time} & \textbf{131 Å} & \textbf{171 Å} & \textbf{304 Å} & \textbf{1600 Å} \\
        \hline
        \multicolumn{6}{c}{\textbf{Eruption Energy (erg)}} \\
        \hline
        M1.1 & 2024-03-23T06:47 & \(2.41 \times 10^{25}\) & \(2.41 \times 10^{26}\) & \(3.92 \times 10^{27}\) & \(8.45 \times 10^{26}\) \\
        M1.1* & 2025-01-07T22:35 & \(3.44 \times 10^{26}\) & \(2.83 \times 10^{27}\) & \(1.73 \times 10^{28}\) & \(2.52 \times 10^{27}\) \\
        M1.9 & 2024-12-21T00:33 & \(1.49 \times 10^{24}\) & \(1.26 \times 10^{25}\) & \(2.05 \times 10^{26}\) & \(1.80 \times 10^{25}\) \\
        M2.0 & 2022-03-02T17:31 & \(5.30 \times 10^{25}\) & \(3.32 \times 10^{26}\) & \(5.28 \times 10^{27}\) & \(1.06 \times 10^{27}\) \\
        M2.7 & 2025-01-24T20:48 & \(3.09 \times 10^{25}\) & \(2.42 \times 10^{26}\) & \(1.91 \times 10^{27}\) & \(4.97 \times 10^{26}\) \\
        M3.8 & 2024-12-19T15:27 & \(8.93 \times 10^{24}\) & \(2.35 \times 10^{25}\) & \(4.23 \times 10^{26}\) & \(2.83 \times 10^{25}\) \\
        X1.0 & 2024-05-12T16:11 & \(2.40 \times 10^{26}\) & \(4.47 \times 10^{26}\) & \(5.86 \times 10^{27}\) & \(1.84 \times 10^{27}\) \\
        X2.0 & 2024-10-31T21:12 & \(1.64 \times 10^{26}\) & \(4.04 \times 10^{26}\) & \(9.31 \times 10^{27}\) & \(1.54 \times 10^{27}\) \\
        X9.0 & 2024-10-03T12:08 & \(2.32 \times 10^{26}\) & \(1.45 \times 10^{27}\) & \(1.54 \times 10^{28}\) & \(5.06 \times 10^{27}\) \\
        \hline
        \multicolumn{6}{c}{\textbf{Ribbon Energy (erg)}} \\
        \hline
        M1.1 & 2024-03-23T06:47 & \(1.19 \times 10^{26}\) & \(4.49 \times 10^{26}\) & \(6.63 \times 10^{27}\) & \(1.30 \times 10^{27}\) \\
        M1.1* & 2025-01-07T22:35 & \(2.36 \times 10^{24}\) & \(1.11 \times 10^{25}\) & \(5.26 \times 10^{26}\) & \(6.44 \times 10^{25}\) \\
        M1.9 & 2024-12-21T00:33 & \(2.27 \times 10^{25}\) & \(1.05 \times 10^{26}\) & \(1.48 \times 10^{27}\) & \(9.23 \times 10^{26}\) \\
        M2.0 & 2022-03-02T17:31 & \(3.07 \times 10^{26}\) & \(4.85 \times 10^{26}\) & \(1.02 \times 10^{28}\) & \(4.48 \times 10^{27}\) \\
        M2.7 & 2025-01-24T20:48 & \(8.27 \times 10^{26}\) & \(7.67 \times 10^{26}\) & \(1.28 \times 10^{28}\) & \(4.97 \times 10^{27}\) \\
        M3.8 & 2024-12-19T15:27 & \(5.15 \times 10^{25}\) & \(2.45 \times 10^{26}\) & \(4.92 \times 10^{27}\) & \(1.95 \times 10^{27}\) \\
        X1.0 & 2024-05-12T16:11 & \(1.54 \times 10^{27}\) & \(6.72 \times 10^{26}\) & \(8.46 \times 10^{27}\) & \(7.29 \times 10^{27}\) \\
        X2.0 & 2024-10-31T21:12 & \(2.87 \times 10^{27}\) & \(5.20 \times 10^{27}\) & \(9.36 \times 10^{28}\) & \(3.49 \times 10^{28}\) \\
        X9.0 & 2024-10-03T12:08 & \(3.05 \times 10^{27}\) & \(5.13 \times 10^{27}\) & \(1.18 \times 10^{29}\) & \(4.79 \times 10^{28}\) \\
        \hline
        \multicolumn{6}{c}{\textbf{Percentage of Total Radiated Energy by Eruption (\%)}} \\
        \hline
        M1.1 & 2024-03-23T06:47 & 17 & 35 & 37 & 39 \\
        M1.1* & 2025-01-07T22:35 & 99 & 100 & 97 & 98 \\
        M1.9 & 2024-12-21T00:33 & 6 & 11 & 12 & 2 \\
        M2.0 & 2022-03-02T17:31 & 15 & 41 & 34 & 19 \\
        M2.7 & 2025-01-24T20:48 & 4 & 24 & 13 & 9 \\
        M3.8 & 2024-12-19T15:27 & 15 & 9 & 8 & 1 \\
        X1.0 & 2024-05-12T16:11 & 13 & 40 & 41 & 20 \\
        X2.0 & 2024-10-31T21:12 & 5 & 7 & 9 & 4 \\
        X9.0 & 2024-10-03T12:08 & 7 & 22* & 12 & 10 \\
        \hline
        \multicolumn{6}{c}{\textbf{Averages}}\\
        \hline
        \textbf{Eruption Energy (erg)} & -- & 
        \(9.43 \times 10^{25} \) & 
        % \(3.93 \times 10^{26} \) & 
        \(2.43 \times 10^{26} \) &
        \(5.29 \times 10^{27} \) & 
        \(1.36 \times 10^{27} \) \\[6pt]
    
        \textbf{Ribbon Energy (erg)} & -- & 
        \(1.10 \times 10^{27}\) & 
        % \(1.63 \times 10^{27}\) & 1.1317918701027036e+27
        \(1.13 \times 10^{27}\) &
        \(3.20 \times 10^{28}\) & 
        \(1.30 \times 10^{28}\) \\[6pt]
        
        \textbf{Eruption Percentage (\%)} & -- & 
        \(10^{+4}_{-4}\) & 
        % \(24^{+13}_{-13}\) & 
        \(24^{+14}_{-14}\) & 
        \(21^{+14}_{-10}\) & 
        \(13^{+6}_{-9}\) \\[6pt]
        
        \textbf{Fraction IQR (\%)} & -- & 
        \(9\) & 
        % \(26\) & 
        \(28\) & 
        \(24\) & 
        \(16\) \\
        \hline
        \end{tabular}
        \caption{Impulsive phase (GOES start to peak time) radiated energies (erg) for eruptions and ribbons with percentage of total (eruption and ribbon) energy radiated by the eruption, in $131\,\textrm{Å}$, $171\,\textrm{Å}$, $304\,\textrm{Å}$ and $1600\,\textrm{Å}$ emission for nine M- and X-class flares. Average radiated energies and percentages of energy radiated by eruption for the events (*excluding M1.1 on 7 January 2025 for all channels, and the X9.0 flare on 3 October 2024 for the $171\,\textrm{Å}$ channel) are also listed.}
        \label{table6.1}
        \end{table}
        
        The masking technique to separate flare ribbon and eruption irradiance contributions was applied to a further six events, providing small-sample insights into the typical contributions of eruptions to irradiance enhancements during M- and X-class flares. Radiated energies over the impulsive phase by ribbons and eruptions for $131\,\textrm{Å}$, $171\,\textrm{Å}$, $304\,\textrm{Å}$ and $1600\,\textrm{Å}$ emission measured by AIA, along with the percentages of energy radiated by the eruptions, are presented in Table \ref{table6.1}. Across the sample, the average percentage of energy radiated by the eruption was greatest for $171\,\textrm{Å}$ emission at $24^{+14}_{-14}\%$, with averages of $10^{+4}_{-4}\%$, $21^{+14}_{-10}\%$ and $13^{+6}_{-9}\%$ for the $131\,\textrm{Å}$, $304\,\textrm{Å}$ and $1600\,\textrm{Å}$ channels, respectively. The plus and minus errors in these percentages represent the difference between the sample mean and the sample's upper and lower quartile values, respectively. These percentages corresponded to absolute radiated energies by eruptions of $2.43\times10^{26}\,\textrm{erg}$, $9.43\times10^{25}\,\textrm{erg}$, $5.29\times10^{27}\,\textrm{erg}$ and $1.36\times10^{27}\,\textrm{erg}$ for the respective channels.
    
        The greatest radiated energy by an eruption was $1.73\times10^{28}\,\textrm{erg}$ in $304\,\textrm{Å}$ emission for the M1.1 flare on 7 January 2025, with the most energy radiated by ribbons being $1.19\times10^{29}\,\textrm{erg}$, also in $304\,\textrm{Å}$ emission, during the X9.0 event. The fraction of total energy radiated by an eruption was also greatest during the M1.1 on 7 January 2025, at $100\%$ for $171\,\textrm{Å}$ emissions, due to the event's ribbons being occulted. Aside from this event, in AIA data, the largest percentage of total radiated energy by an eruption was $41\%$ for $304\,\textrm{Å}$ emissions during the X1.0 flare. However, HRI$_{\mathrm{Lya}}$ measured a larger $55\%$ during the M2.0 event, potentially due to the line's large optical depth in the chromosphere, with erupted material perhaps more efficiently radiating away energy in the line due to lower opacity in the corona \citep{Fontenla_1988}.
        
        \begin{table}[ht]
        \begin{tabular}{llcc}
        \hline
        \multicolumn{4}{c}{\textbf{Eruption Energy (erg)}} \\
        \hline
        \textbf{Class} & \textbf{Start Time} & \textbf{304 Å} & \textbf{1600 Å} \\
        \hline
        M1.1 & 2024-03-23T06:47 & \(1.31 \times 10^{28}\) & \(1.69 \times 10^{27}\) \\
        M1.1* & 2025-01-07T22:35 & \(1.93 \times 10^{28}\) & \(3.24 \times 10^{27}\) \\
        M1.9 & 2024-12-21T00:33 & \(1.98 \times 10^{27}\) & \(1.78 \times 10^{26}\) \\
        M2.0 & 2022-03-02T17:31 & \(1.28 \times 10^{28}\) & \(2.69 \times 10^{27}\) \\
        M2.7 & 2025-01-24T20:48 & \(3.98 \times 10^{27}\) & \(8.98 \times 10^{26}\) \\
        M3.8 & 2024-12-19T15:27 & \(3.63 \times 10^{27}\) & \(2.72 \times 10^{26}\) \\
        X1.0 & 2024-05-12T16:11 & \(1.49 \times 10^{28}\) & \(3.97 \times 10^{27}\) \\
        X2.0 & 2024-10-31T21:12 & \(2.31 \times 10^{28}\) & \(3.34 \times 10^{27}\) \\
        X9.0 & 2024-10-03T12:08 & \(4.56 \times 10^{28}\) & \(1.20 \times 10^{28}\) \\
        \hline
        \multicolumn{4}{c}{\textbf{Ribbon Energy (erg)}} \\
        \hline
        M1.1 & 2024-03-23T06:47 & \(1.03 \times 10^{28}\) & \(1.98 \times 10^{27}\) \\
        M1.1* & 2025-01-07T22:35 & \(1.91 \times 10^{27}\) & \(1.33 \times 10^{26}\) \\
        M1.9 & 2024-12-21T00:33 & \(6.37 \times 10^{27}\) & \(2.15 \times 10^{27}\) \\
        M2.0 & 2022-03-02T17:31 & \(2.53 \times 10^{28}\) & \(9.22 \times 10^{27}\) \\
        M2.7 & 2025-01-24T20:48 & \(2.26 \times 10^{28}\) & \(7.62 \times 10^{27}\) \\
        M3.8 & 2024-12-19T15:27 & \(1.07 \times 10^{28}\) & \(4.11 \times 10^{27}\) \\
        X1.0 & 2024-05-12T16:11 & \(2.25 \times 10^{28}\) & \(1.15 \times 10^{28}\) \\
        X2.0 & 2024-10-31T21:12 & \(1.59 \times 10^{29}\) & \(4.83 \times 10^{28}\) \\
        X9.0 & 2024-10-03T12:08 & \(3.70 \times 10^{29}\) & \(1.01 \times 10^{29}\) \\
        \hline
        \multicolumn{4}{c}{\textbf{Percentage of Total Radiated Energy by Eruption (\%)}} \\
        \hline
        M1.1 & 2024-03-23T06:47 & 56 & 46 \\
        M1.1* & 2025-01-07T22:35 & 91 & 96 \\
        M1.9 & 2024-12-21T00:33 & 24 & 8 \\
        M2.0 & 2022-03-02T17:31 & 34 & 23 \\
        M2.7 & 2025-01-24T20:48 & 15 & 11 \\
        M3.8 & 2024-12-19T15:27 & 25 & 6 \\
        X1.0 & 2024-05-12T16:11 & 40 & 26 \\
        X2.0 & 2024-10-31T21:12 & 13 & 6 \\
        X9.0 & 2024-10-03T12:08 & 11 & 11 \\
        \hline
        \multicolumn{4}{c}{\textbf{Averages}} \\
        \hline
        \multicolumn{2}{l}{\textbf{Eruption Energy (erg)}} &
          \(1.49 \times 10^{28}\) &
          \(3.13 \times 10^{27}\) \\
        \multicolumn{2}{l}{\textbf{Ribbon Energy (erg)}} &
          \(7.84 \times 10^{28}\) &
          \(2.32 \times 10^{28}\) \\
        \multicolumn{2}{l}{\textbf{Eruption Percentage (\%)}} &
          \(27^{+8}_{-13}\) &
          \(17^{+6}_{-10}\) \\
        \multicolumn{2}{l}{\textbf{Fraction IQR (\%)}} &
          \(21\) &
          \(16\) \\
        \hline
        \end{tabular}
        \caption{Radiated energies (erg) over GOES flare period for eruptions and ribbons with percentage of total (eruption and ribbon) energy radiated by the eruption, in $304\,\textrm{Å}$ and $1600\,\textrm{Å}$ emission for nine M- and X-class flares. Average radiated energies and percentages of energy radiated by eruption for the events (*excluding M1.1 on 7 January 2025) are also listed.}
        \label{table6.2}
        \end{table}
    
        Gradual phase contributions were not considered for $131\,\textrm{Å}$ and $171\,\textrm{Å}$ emissions due to potential contamination from flare loop emission and coronal dimming. However, Table \ref{table6.2} lists eruption and ribbon radiated energies over the entire GOES flare durations for $304\,\textrm{Å}$ and $1600\,\textrm{Å}$ emissions. The averaged energies for both eruptions and ribbons are expectedly greater than the impulsive phase values presented in Table \ref{table6.1}. The percentage of energy radiated by the eruption is also greater over the entire flare duration than the impulsive phase alone for both channels, suggesting that eruption emission is relatively stronger in the gradual phase. However, this result has weak significance due to the small sample size.

        Table \ref{table6.3} shows the timing of peak eruption and ribbon emission for the $304\,\textrm{Å}$ and $1600\,\textrm{Å}$ channels. The other channels are again excluded due to flare loop contributions and coronal dimming. It is seen that the eruption intensity peaks after or at the same time as the ribbons for seven of eight events in the $304\,\textrm{Å}$ emission and for all eight events in $1600\textrm{Å}$, omitting the M1.1 event, which had occulted ribbons. This may suggest that a heating mechanism other than the deposition of nonthermal electron energy drives the enhancement of erupted material. Other possible mechanisms may include conduction of energy from the ribbons or heating through Ohmic dissipation of reconnection-driven currents. The largest time delay between peak ribbon and eruption emission occurred during the X2.0 event, which had its eruption emission peak over 30 minutes after its ribbons. For the M1.1 flare on 7 January 2025, the flare ribbons were not visible from AIA. However, EUVI observations show that the event's eruption peaks in brightness around 10 minutes before the ribbons. This indicates that while eruption and ribbon emission typically peak close together in time, peak eruption emission can occur tens of minutes either before or after the ribbon peak.
        
        \begin{table}[ht]
        \begin{tabular}{cccccc}
        \hline
        \textbf{Class} & \textbf{Date} & \textbf{\shortstack{Start\\Time}} & \textbf{\shortstack{Eruption\\Peak Time}} & \textbf{\shortstack{Ribbon\\Peak Time}} & \textbf{\shortstack{\(\Delta \textrm{t}\)\,(s)\\(Erupt-Rib)}}\\
        \hline
        \multicolumn{6}{c}{$304\,\textrm{Å}$} \\
        \hline
        M1.1 & 2024-03-23 & 06:47:00 & 06:57:05 & 06:54:41 & 144 \\
        M1.1* & 2025-01-07 & 22:35:00 & 22:44:05 & N/A & N/A \\
        M1.9 & 2024-12-21 & 00:33:00 & 00:38:29 & 00:37:53 & 36 \\
        M2.0 & 2022-03-02 & 17:31:00 & 17:38:29 & 17:37:53 & 36 \\
        M2.7 & 2025-01-24 & 20:48:00 & 21:04:41 & 20:56:17 & 504 \\
        M3.8 & 2024-12-19 & 15:27:00 & 15:35:53 & 15:33:53 & 120 \\ 
        X1.0 & 2024-05-12 & 16:11:00 & 16:21:53 & 16:19:29 & 144 \\
        X2.0 & 2024-10-31 & 21:12:00 & 21:49:53 & 21:16:55 & 1977 \\
        X9.0 & 2024-10-03 & 12:08:00 & 12:17:29 & 12:17:44 & -15 \\
        \hline
        \multicolumn{6}{c}{$1600\,\textrm{Å}$} \\
        \hline
        M1.1 & 2024-03-23 & 06:47:00 & 06:55:02 & 06:55:02 & 0  \\
        M1.1* & 2025-01-07 & 22:35:00 & 22:43:50 & N/A & N/A  \\
        M1.9 & 2024-12-21 & 00:33:00 & 00:38:38 & 00:37:50 & 48  \\
        M2.0 & 2022-03-02 & 17:31:00 & 17:38:14 & 17:37:50 & 24  \\
        M2.7 & 2025-01-24 & 20:48:00 & 21:04:38 & 20:55:50 & 528 \\
        M3.8 & 2024-12-19 & 15:27:00 & 15:35:50 & 15:33:50 & 120 \\
        X1.0 & 2024-05-12 & 16:11:00 & 16:19:26 & 16:19:26 & 0  \\
        X2.0 & 2024-10-31 & 21:12:00 & 21:49:02 & 21:15:50 & 1992 \\
        X9.0 & 2024-10-03 & 12:08:00 & 12:17:26 & 12:16:38 & 48  \\
        \hline
        \end{tabular}
        \caption{Timing of peak AIA $304\,\textrm{Å}$ and $1600\,\textrm{Å}$ emission from eruption and ribbons for nine M- and X-class flares, with time difference between eruption and ribbon peak. *Timing differences not noted for M1.1 flare as ribbons were occulted.}
        \label{table6.3}
        \end{table}
        
        %SHOULD TABULATE VALUES FOR THIS?
        Examining EUVS-B \Ly data for each event reveals an average peak flare percentage enhancement above background of $3.16\%$, with the first data point after the GOES start time considered as a background. Such a value is consistent with typical values for M- and X-class events, suggesting that flare-associated eruptions may not drive significantly larger enhancements \citep{Milligan_2020,Milligan_2021}. Few events in the sample showed appreciable \Ly enhancements in LYRA observations, the largest enhancement being during the X9.0 event at $\sim1\%$, considerably less than the $13\%$ measured by EUVS-B for the same event. This disparity may have been due to contamination of the LYRA bandpass by continuum emission \citep{Greatorex_2024}.
        
        \subsection{Eruption Detection in Photometric Observations}
\begin{figure}[ht!] 
            \centerline{
                    \includegraphics[width =0.6 \paperwidth,clip=]{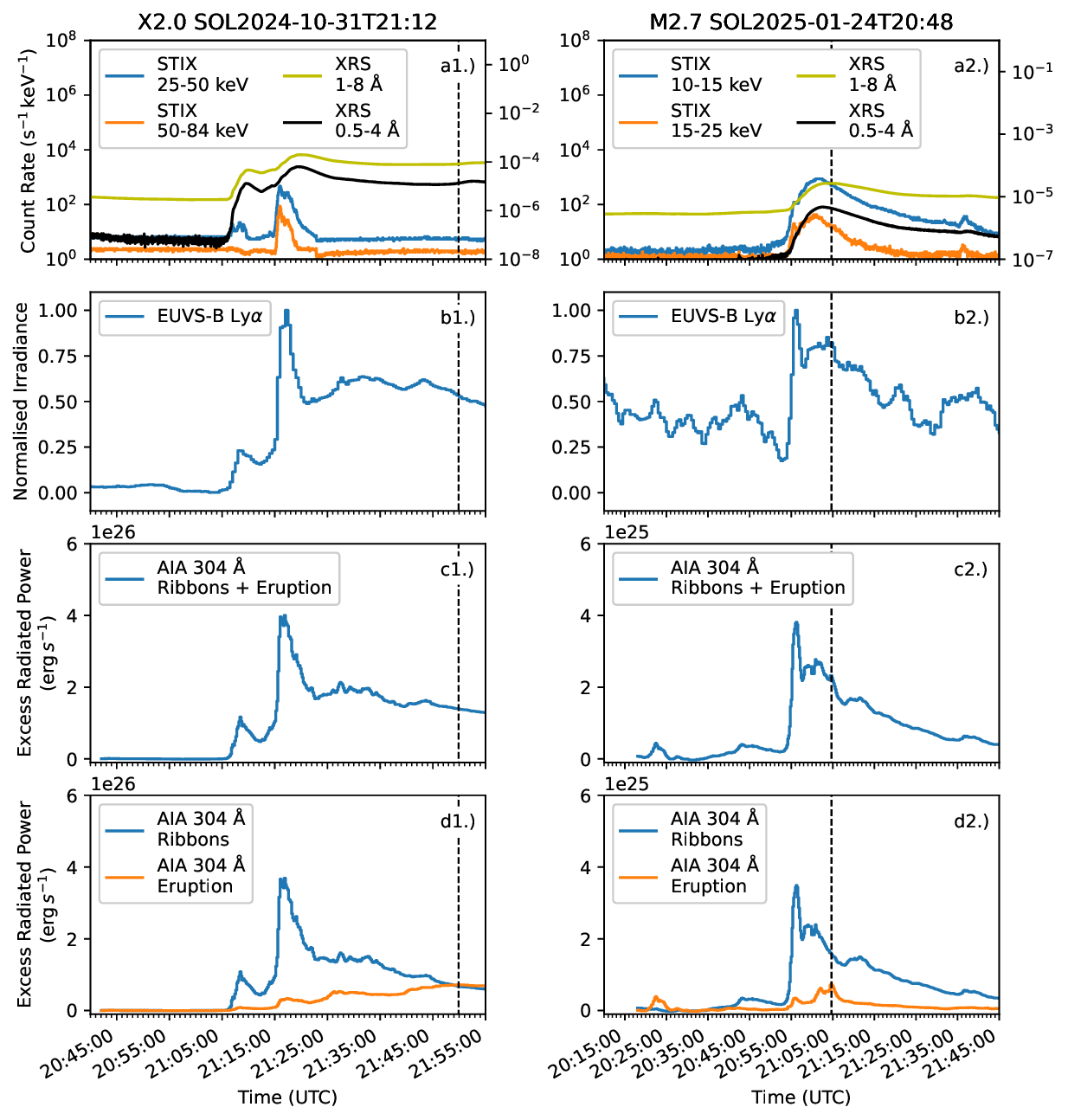}
                  }
        \caption{X-ray and UV emissions during an X2.0 (left) and M2.7 flare (right). Panels a1.) and a2.) show X-ray lightcurves from STIX and XRS. Excess \Ly emission measured by EUVS-B is shown in panels b1.) and b2.). Radiated energy for $304\,\textrm{Å}$ against time for the combined ribbons and erupted material is plotted in panels c1.) and c2.). Divided ribbon and eruption radiated energy is plotted in panels d1.) and d2.). Dashed black lines in each panel represent the time of peak emission from each event's eruption.}
        \label{figure6.11}
        \end{figure}

        Figure \ref{figure6.11} shows lightcurves for the X2.0 and M2.7 events, which both had eruption emission peak more than 5 minutes after the flare ribbons. For the X2.0 event, the peak of $304\,\textrm{Å}$ eruption emission occurs with no associated HXR emission as observed by STIX, providing a potential signature of eruption emission in Sun-as-a-star data. However, the eruption and ribbon contributions for this event cannot be delineated in photometric \Ly observations, as the increase in eruption emission does not show a sharp peak. 

        In contrast, during the M2.7 event, peak emission of erupted material in $304\,\textrm{Å}$ occurred between peaks in ribbon emission. This later peak in ribbon emission does not correlate with a burst in HXRs, potentially suggesting this secondary peak was driven by thermal conduction of energy from the flare loops to the ribbons. Hence, comparison of photometric enhancements to HXR data may not allow eruption and ribbon contributions to be reliably distinguished during flares. 

    \section{Discussion \& Conclusions}
    \label{section4}
    In this work, we provide an analysis of nine flares with associated bright eruptions. Three events were studied in detail, including an M2.0 event with an eruption seen on disk, an X9.0 event with a relatively less dynamic eruption, and an eruptive M1.1 event with its ribbons fully occulted from Earth's viewpoint. Basic statistics on the radiative contributions of eruptions are provided from the full sample, using AIA data. Over the impulsive phase, the average percentage of total flare energy radiated by the eruptions was between $10\%$ in the hot $131\,\textrm{Å}$ channel and $24\%$ in the relatively cooler $171\,\textrm{Å}$ channel, with averages of $21\%$ and $13\%$ for the $304\,\textrm{Å}$ and $1600\,\textrm{Å}$ channels, respectively. This demonstrates that erupted material can radiate a substantial fraction of the total radiated energy in UV wavelengths during flares\comment. However, it is noted that many of the flares in an initial sample of 82 events did not display a bright eruption, suggesting that this phenomenon is rare\commentend. As these eruptions can move at hundreds of $\textrm{km\,s}^{-1}$ \citep[e.g.][]{Mccauley_2015}, they may drive blueshifts in Sun-as-a-star flare spectra, which could otherwise be interpreted as a signature of chromospheric evaporation \citep{Batchelor_1991,Majury_2025_1}. This eruption-driven effect may also drive similar shifts in observations of stellar flares \citep{Gunn_1994,Berdyugina_1999,Wang_2024}. Identification of eruptive events via coronal dimming may help clarify whether observed shifts are likely due to eruptions or evaporation in spatially integrated data \citep{Veronig_2021}. 
    
    Synthesised flare profiles from contemporary flare simulations such as RADYN \citep{Allred_2005,Allred_2015}, HYDRO2GEN \citep{Druett_2018,Druett_2019}, and FLARIX \citep{Varady_2010,Heinzel_2016} may provide an avenue to disentangle ribbon and eruption flare contributions to observed flare irradiance and spectral variability, as these simulations typically do not include eruptions. Furthermore, the inclusion of eruptions in contemporary flare simulations may help clarify the physical mechanisms driving observed emissions. This may be possible through further development of existing MHD simulations of solar eruptive events \citep[e.g.][]{Ruan_2020,Jenkins_2021,Jenkins_2022,Jenkins_2023,Druett_2024}. 
    
    While the eruptions in this sample were found to radiate a substantial fraction of the radiated energy, the total excess radiated energies were comparable to those seen in confined events \citep{Greatorex_2023}. Thus, it remains unclear whether eruptive events tend to radiate a larger fraction of energy released during flares than events without eruptions. Comparative analysis of the energy release in eruptive and confined flares may further clarify whether eruptive events radiate released flare energy more efficiently.
    
    For three of the nine events, HXR images were generated, providing insight into the heating mechanism driving enhanced emission from their associated eruptions. For the M2.0 event on 2 March 2022, we find only limited overlap of HXR emission and erupted material, despite the material radiating a large fraction of the excess radiated energy. Hence, it is unclear whether local nonthermal heating was primarily responsible for the enhancement of the erupted material. The X9.0 event showed no overlap of HXR sources with erupted material. This suggests the enhancement of the material was not strongly driven by nonthermal heating. It may be possible that there was substantial HXR emission from the erupted material that could not have been easily detected due to the dynamic range of STIX and HXI. However, we found no HXR emission during the occulted M1.1 event, further suggesting the heating mechanism for the observed bright eruptions could not have been nonthermal particle heating alone. 
    
    For the occulted M1.1 event, it seems unlikely that its eruption was heated through thermal conduction of energy from the flare ribbons, as the eruption peaked in brightness before the ribbons. Other heating mechanisms, such as Ohmic heating or heating by MHD waves, may have instead produced the observed brightening \citep{Xue_2016,Reeves_2019}. However, further observations and modelling are required to confirm this. The emission from the occulted M1.1 event somewhat resembles the bright eruption observed in \citet{Hayes_2024}, although, the event detailed in their study had thermal and nonthermal X-ray emission cospatial with erupted material, with the authors suggesting the heating mechanism to be deposition of energy by upwardly accelerated electrons with an inferred nonthermal energy of $\sim2\times10^{28}\,\textrm{erg}$. The total energy radiated by the occulted M1.1 event's eruption over the GOES period in $304\,\textrm{Å}$ emission was $1.93\times10^{28}\,\textrm{erg}$, which is on the order of the average energy radiated by ribbons in the sample of $7.84\times 10^{28}\,\textrm{erg}$, despite the event having the joint smallest GOES class in the sample. As He~\textsc{ii} $304\,\textrm{Å}$ emission strongly contributes to flare heating of the Earth's ionosphere, similarly occulted events may produce unexpectedly strong ionospheric enhancements for their GOES class \citep{Watanabe_2021,Ohare_2025}. If the flare loops are fully occulted but the bright erupted material is still visible, an ionospheric disturbance without a GOES classified flare may be observed. A search for such an event in AIA observations and subsequent analysis of total ionospheric electron content may further clarify the importance of this phenomenon to space weather.

    Further insights into the radiative properties of eruptions may be provided via analysis of imaging spectrometry observations from existing instruments such as the EUV Imaging Spectrometer on Hinode \citep[Hinode/EIS;][]{Culhane_2007} and the Interface Region Imaging Spectrograph \citep[IRIS;][]{Depontieu_2014}, along with upcoming instruments such as the EUV High-throughput Spectroscopic Telescope on Solar-C \citep[Solar-C/EUVST;][]{Shimizu_2019}. Such studies may provide further insight into the contribution of eruption emission to observed line profile shifts in Sun-as-a-star observations, and guide the interpretation of observations of line shifts in stellar flares. Analysis of spectral observations may also help constrain the heating mechanism of erupted material through measurement of nonthermal broadening \citep{Russell_2024}. 
    
    In summary, this work finds that a substantial amount of flare excess radiated energy in UV emissions can originate from erupted material. We demonstrate that the brightening of erupted material is unlikely to be ubiquitously driven by nonthermal particle heating. Our work has implications for the interpretation of solar and stellar flare observations, with both eruptions and chromospheric evaporation potentially driving line shifts. Future observational and modelling work may provide further insights into the relative importance of these bright eruptions in observed spectral variability and elucidate the typical heating source responsible for the observed brightening.
       
\begin{fundinginformation}
L.H.M. acknowledges support from the Department for the Economy (DfE) Northern Ireland postgraduate studentship scheme, and from BELSPO through an EUI and SWAP/LYRA Guest Investigator grant. R.O.M. acknowledges support from STFC grants ST/W001144/1 and ST/X000923/1.
M.D. and D.C.T. acknowledge support from BELSPO in the framework of the ESA-PRODEX program, PEA 4000145189 and 4000136681.
LYRA is a project of the Centre Spatial de Liège, the Physikalisch-Meteorologisches Observatorium Davos and the Royal Observatory of Belgium, funded by the Belgian Federal Science Policy Office (BELSPO) and by the Swiss Bundesamt für Bildung und Wissenschaft.
Solar Orbiter is a space mission of international collaboration between ESA and NASA, operated by ESA. The EUI instrument was built by CSL, IAS, MPS, MSSL/UCL, PMOD/WRC, ROB, LCF/IO with funding from the Belgian Federal Science Policy Office (BELSPO/PRODEX PEA 4000112292 and 4000134088); the Centre National d’Etudes Spatiales (CNES); the UK Space Agency (UKSA); the Bundesministerium für Wirtschaft und Energie (BMWi) through the Deutsches Zentrum für Luft- und Raumfahrt (DLR); and the Swiss Space Office (SSO).
\end{fundinginformation}

\begin{dataavailability}
The CACTUS CME catalogue is accessible at https://www.sidc.be/cactus/
\end{dataavailability}

\bibliographystyle{spr-mp-sola}
\bibliography{bibliography.bib} 

@INPROCEEDINGS{Shimizu_2019,
       author = {{Shimizu}, Toshifumi and {Imada}, Shinsuke and {Kawate}, Tomoko and {Ichimoto}, Kiyoshi and {Suematsu}, Yoshinori and {Hara}, Hirohisa and {Katsukawa}, Yukio and {Kubo}, Masahito and {Toriumi}, Shin and {Watanabe}, Tetsuya and {Yokoyama}, Takaaki and {Korendyke}, Clarence M. and {Warren}, Harry P. and {Tarbell}, Ted and {De Pontieu}, Bart and {Teriaca}, Luca and {Sch{\"u}hle}, Udo H. and {Solanki}, Sami and {Harra}, Louise K. and {Matthews}, Sarah and {Fludra}, A. and {Auch{\`e}re}, F. and {Andretta}, V. and {Naletto}, G. and {Zhukov}, A.},
        title = "{The Solar-C\_EUVST mission}",
    booktitle = {UV, X-Ray, and Gamma-Ray Space Instrumentation for Astronomy XXI},
         year = 2019,
       editor = {{Siegmund}, Oswald H.},
       series = {Society of Photo-Optical Instrumentation Engineers (SPIE) Conference Series},
       volume = {11118},
        month = sep,
          eid = {1111807},
        pages = {1111807},
          doi = {10.1117/12.2528240},
       adsurl = {https://ui.adsabs.harvard.edu/abs/2019SPIE11118E..07S}
}

@ARTICLE{Majury_2025,
       author = {{Majury}, Luke and {Milligan}, Ryan and {Butler}, Elizabeth and {Greatorex}, Harry and {Kazachenko}, Maria},
        title = "{Spectral Irradiance Variability in Lyman-Alpha Emission During Solar Flares}",
      journal = {arXiv e-prints},
         year = 2025,
        month = apr,
          eid = {arXiv:2504.17667},
        pages = {arXiv:2504.17667},
archivePrefix = {arXiv},
       eprint = {2504.17667},
 primaryClass = {astro-ph.SR},
       adsurl = {https://ui.adsabs.harvard.edu/abs/2025arXiv250417667M}
}

@ARTICLE{Hurlburt_2015,
       author = {{Hurlburt}, N.},
        title = "{Automated detection of solar eruptions}",
      journal = {Journal of Space Weather and Space Climate},
         year = 2015,
        month = dec,
       volume = {5},
          eid = {A39},
        pages = {A39},
          doi = {10.1051/swsc/2015042},
archivePrefix = {arXiv},
       eprint = {1504.03395},
 primaryClass = {astro-ph.SR},
       adsurl = {https://ui.adsabs.harvard.edu/abs/2015JSWSC...5A..39H}
}

@ARTICLE{Hurlburt_2012,
       author = {{Hurlburt}, N. and {Cheung}, M. and {Schrijver}, C. and {Chang}, L. and {Freeland}, S. and {Green}, S. and {Heck}, C. and {Jaffey}, A. and {Kobashi}, A. and {Schiff}, D. and {Serafin}, J. and {Seguin}, R. and {Slater}, G. and {Somani}, A. and {Timmons}, R.},
        title = "{Heliophysics Event Knowledgebase for the Solar Dynamics Observatory (SDO) and Beyond}",
      journal = {\solphys},
         year = 2012,
        month = jan,
       volume = {275},
       number = {1-2},
        pages = {67-78},
          doi = {10.1007/s11207-010-9624-2},
archivePrefix = {arXiv},
       eprint = {1008.1291},
 primaryClass = {astro-ph.IM},
       adsurl = {https://ui.adsabs.harvard.edu/abs/2012SoPh..275...67H}
}

@ARTICLE{Muller_2017,
       author = {{M{\"u}ller}, D. and {Nicula}, B. and {Felix}, S. and {Verstringe}, F. and {Bourgoignie}, B. and {Csillaghy}, A. and {Berghmans}, D. and {Jiggens}, P. and {Garc{\'\i}a-Ortiz}, J.~P. and {Ireland}, J. and {Zahniy}, S. and {Fleck}, B.},
        title = "{JHelioviewer. Time-dependent 3D visualisation of solar and heliospheric data}",
      journal = {\aap},
         year = 2017,
        month = sep,
       volume = {606},
          eid = {A10},
        pages = {A10},
          doi = {10.1051/0004-6361/201730893},
archivePrefix = {arXiv},
       eprint = {1705.07628},
 primaryClass = {astro-ph.SR},
       adsurl = {https://ui.adsabs.harvard.edu/abs/2017A&A...606A..10M}
}

@ARTICLE{Berghmans_2023,
       author = {{Berghmans}, D. and {Antolin}, P. and {Auch{\`e}re}, F. and {Aznar Cuadrado}, R. and {Barczynski}, K. and {Chitta}, L.~P. and {Gissot}, S. and {Harra}, L. and {Huang}, Z. and {Janvier}, M. and {Kraaikamp}, E. and {Long}, D.~M. and {Mandal}, S. and {Mierla}, M. and {Parenti}, S. and {Peter}, H. and {Rodriguez}, L. and {Sch{\"u}hle}, U. and {Smith}, P.~J. and {Solanki}, S.~K. and {Stegen}, K. and {Teriaca}, L. and {Verbeeck}, C. and {West}, M.~J. and {Zhukov}, A.~N. and {Appourchaux}, T. and {Aulanier}, G. and {Buchlin}, E. and {Delmotte}, F. and {Gilles}, J.~M. and {Haberreiter}, M. and {Halain}, J. -P. and {Heerlein}, K. and {Hochedez}, J. -F. and {Gyo}, M. and {Poedts}, S. and {Renotte}, E. and {Rochus}, P.},
        title = "{First perihelion of EUI on the Solar Orbiter mission}",
      journal = {\aap},
         year = 2023,
        month = jul,
       volume = {675},
          eid = {A110},
        pages = {A110},
          doi = {10.1051/0004-6361/202245586},
archivePrefix = {arXiv},
       eprint = {2301.05616},
 primaryClass = {astro-ph.SR},
       adsurl = {https://ui.adsabs.harvard.edu/abs/2023A&A...675A.110B}
}

@ARTICLE{Krucker_2020,
       author = {{Krucker}, S{\"a}m and {Hurford}, G.~J. and {Grimm}, O. and {K{\"o}gl}, S. and {Gr{\"o}belbauer}, H. -P. and {Etesi}, L. and {Casadei}, D. and {Csillaghy}, A. and {Benz}, A.~O. and {Arnold}, N.~G. and {Molendini}, F. and {Orleanski}, P. and {Schori}, D. and {Xiao}, H. and {Kuhar}, M. and {Hochmuth}, N. and {Felix}, S. and {Schramka}, F. and {Marcin}, S. and {Kobler}, S. and {Iseli}, L. and {Dreier}, M. and {Wiehl}, H.~J. and {Kleint}, L. and {Battaglia}, M. and {Lastufka}, E. and {Sathiapal}, H. and {Lapadula}, K. and {Bednarzik}, M. and {Birrer}, G. and {Stutz}, St. and {Wild}, Ch. and {Marone}, F. and {Skup}, K.~R. and {Cichocki}, A. and {Ber}, K. and {Rutkowski}, K. and {Bujwan}, W. and {Juchnikowski}, G. and {Winkler}, M. and {Darmetko}, M. and {Michalska}, M. and {Seweryn}, K. and {Bia{\l}ek}, A. and {Osica}, P. and {Sylwester}, J. and {Kowalinski}, M. and {{\'S}cis{\l}owski}, D. and {Siarkowski}, M. and {St{\k{e}}{\'s}licki}, M. and {Mrozek}, T. and {Podg{\'o}rski}, P. and {Meuris}, A. and {Limousin}, O. and {Gevin}, O. and {Le Mer}, I. and {Brun}, S. and {Strugarek}, A. and {Vilmer}, N. and {Musset}, S. and {Maksimovi{\'c}}, M. and {F{\'a}rn{\'\i}k}, F. and {Koz{\'a}{\v{c}}ek}, Z. and {Ka{\v{s}}parov{\'a}}, J. and {Mann}, G. and {{\"O}nel}, H. and {Warmuth}, A. and {Rendtel}, J. and {Anderson}, J. and {Bauer}, S. and {Dionies}, F. and {Paschke}, J. and {Pl{\"u}schke}, D. and {Woche}, M. and {Schuller}, F. and {Veronig}, A.~M. and {Dickson}, E.~C.~M. and {Gallagher}, P.~T. and {Maloney}, S.~A. and {Bloomfield}, D.~S. and {Piana}, M. and {Massone}, A.~M. and {Benvenuto}, F. and {Massa}, P. and {Schwartz}, R.~A. and {Dennis}, B.~R. and {van Beek}, H.~F. and {Rodr{\'\i}guez-Pacheco}, J. and {Lin}, R.~P.},
        title = "{The Spectrometer/Telescope for Imaging X-rays (STIX)}",
      journal = {\aap},
         year = 2020,
        month = oct,
       volume = {642},
          eid = {A15},
        pages = {A15},
          doi = {10.1051/0004-6361/201937362},
       adsurl = {https://ui.adsabs.harvard.edu/abs/2020A&A...642A..15K}
}

@ARTICLE{Massa_2020,
       author = {{Massa}, Paolo and {Schwartz}, Richard and {Tolbert}, A. Kim and {Massone}, Anna Maria and {Dennis}, Brian R. and {Piana}, Michele and {Benvenuto}, Federico},
        title = "{MEM\_GE: A New Maximum Entropy Method for Image Reconstruction from Solar X-Ray Visibilities}",
      journal = {\apj},
         year = 2020,
        month = may,
       volume = {894},
       number = {1},
          eid = {46},
        pages = {46},
          doi = {10.3847/1538-4357/ab8637},
archivePrefix = {arXiv},
       eprint = {2002.07921},
 primaryClass = {astro-ph.SR},
       adsurl = {https://ui.adsabs.harvard.edu/abs/2020ApJ...894...46M}
}

@ARTICLE{Hogbom_1974,
       author = {{H{\"o}gbom}, J.~A.},
        title = "{Aperture Synthesis with a Non-Regular Distribution of Interferometer Baselines}",
      journal = {\aaps},
         year = 1974,
        month = jun,
       volume = {15},
        pages = {417},
       adsurl = {https://ui.adsabs.harvard.edu/abs/1974A&AS...15..417H}
}

@ARTICLE{Wauters_2022,
       author = {{Wauters}, L. and {Dominique}, M. and {Milligan}, R. and {Dammasch}, I.~E. and {Kretzschmar}, M. and {Machol}, J.},
        title = "{Observation of a Flare and Filament Eruption in Lyman-{\ensuremath{\alpha}} on 8 September 2011 by the PRoject for OnBoard Autonomy/Large Yield Radiometer (PROBA2/LYRA)}",
      journal = {\solphys},
         year = 2022,
        month = mar,
       volume = {297},
       number = {3},
          eid = {36},
        pages = {36},
          doi = {10.1007/s11207-022-01963-0},
       adsurl = {https://ui.adsabs.harvard.edu/abs/2022SoPh..297...36W}
}

@ARTICLE{Li_2025,
       author = {{Li}, Dong},
        title = "{Localizing short-period pulsations in hard X-rays and {\ensuremath{\gamma}}-rays during an X9.0 flare}",
      journal = {\aap},
         year = 2025,
        month = mar,
       volume = {695},
          eid = {L4},
        pages = {L4},
          doi = {10.1051/0004-6361/202453613},
archivePrefix = {arXiv},
       eprint = {2502.14262},
 primaryClass = {astro-ph.SR},
       adsurl = {https://ui.adsabs.harvard.edu/abs/2025A&A...695L...4L}
}

@ARTICLE{Heyvaerts_1977,
       author = {{Heyvaerts}, J. and {Priest}, E.~R. and {Rust}, D.~M.},
        title = "{An emerging flux model for the solar phenomenon.}",
      journal = {\apj},
         year = 1977,
        month = aug,
       volume = {216},
        pages = {123-137},
          doi = {10.1086/155453},
       adsurl = {https://ui.adsabs.harvard.edu/abs/1977ApJ...216..123H}
}

@ARTICLE{Antiochos_1999,
       author = {{Antiochos}, S.~K. and {DeVore}, C.~R. and {Klimchuk}, J.~A.},
        title = "{A Model for Solar Coronal Mass Ejections}",
      journal = {\apj},
         year = 1999,
        month = jan,
       volume = {510},
       number = {1},
        pages = {485-493},
          doi = {10.1086/306563},
archivePrefix = {arXiv},
       eprint = {astro-ph/9807220},
 primaryClass = {astro-ph},
       adsurl = {https://ui.adsabs.harvard.edu/abs/1999ApJ...510..485A}
}

@ARTICLE{Moore_2001,
       author = {{Moore}, Ronald L. and {Sterling}, Alphonse C. and {Hudson}, Hugh S. and {Lemen}, James R.},
        title = "{Onset of the Magnetic Explosion in Solar Flares and Coronal Mass Ejections}",
      journal = {\apj},
         year = 2001,
        month = may,
       volume = {552},
       number = {2},
        pages = {833-848},
          doi = {10.1086/320559},
       adsurl = {https://ui.adsabs.harvard.edu/abs/2001ApJ...552..833M}
}

@ARTICLE{Woods_2006,
       author = {{Woods}, Thomas N. and {Kopp}, Greg and {Chamberlin}, Phillip C.},
        title = "{Contributions of the solar ultraviolet irradiance to the total solar irradiance during large flares}",
      journal = {Journal of Geophysical Research (Space Physics)},
         year = 2006,
        month = oct,
       volume = {111},
       number = {A10},
          eid = {A10S14},
        pages = {A10S14},
          doi = {10.1029/2005JA011507},
       adsurl = {https://ui.adsabs.harvard.edu/abs/2006JGRA..11110S14W}
}

@ARTICLE{Milligan_2014,
       author = {{Milligan}, Ryan O. and {Kerr}, Graham S. and {Dennis}, Brian R. and {Hudson}, Hugh S. and {Fletcher}, Lyndsay and {Allred}, Joel C. and {Chamberlin}, Phillip C. and {Ireland}, Jack and {Mathioudakis}, Mihalis and {Keenan}, Francis P.},
        title = "{The Radiated Energy Budget of Chromospheric Plasma in a Major Solar Flare Deduced from Multi-wavelength Observations}",
      journal = {\apj},
         year = 2014,
        month = oct,
       volume = {793},
       number = {2},
          eid = {70},
        pages = {70},
          doi = {10.1088/0004-637X/793/2/70},
archivePrefix = {arXiv},
       eprint = {1406.7657},
 primaryClass = {astro-ph.SR},
       adsurl = {https://ui.adsabs.harvard.edu/abs/2014ApJ...793...70M}
}

@ARTICLE{Curto_2020,
       author = {{Curto}, Juan Jos{\'e}},
        title = "{Geomagnetic solar flare effects: a review}",
      journal = {Journal of Space Weather and Space Climate},
         year = 2020,
        month = may,
       volume = {10},
          eid = {27},
        pages = {27},
          doi = {10.1051/swsc/2020027},
       adsurl = {https://ui.adsabs.harvard.edu/abs/2020JSWSC..10...27C}
}

@BOOK{Mitra_1974,
       author = {{Mitra}, A.~P.},
        title = "{Ionospheric effects of solar flares}",
         year = 1974,
       volume = {46},
          doi = {10.1007/978-94-010-2231-6},
       adsurl = {https://ui.adsabs.harvard.edu/abs/1974ASSL...46.....M}
}

@ARTICLE{Rubiodacosta_2009,
       author = {{Rubio da Costa}, F. and {Fletcher}, L. and {Labrosse}, N. and {Zuccarello}, F.},
        title = "{Observations of a solar flare and filament eruption in Lyman {\ensuremath{\alpha}} and X-rays}",
      journal = {\aap},
         year = 2009,
        month = nov,
       volume = {507},
       number = {2},
        pages = {1005-1014},
          doi = {10.1051/0004-6361/200912651},
archivePrefix = {arXiv},
       eprint = {0909.4705},
 primaryClass = {astro-ph.SR},
       adsurl = {https://ui.adsabs.harvard.edu/abs/2009A&A...507.1005R}
}

@ARTICLE{Batchelor_1991,
       author = {{Batchelor}, D.~A. and {Hindsley}, K.~P.},
        title = "{X-ray observations of filament eruption in the 1980 May 21 flare}",
      journal = {\solphys},
         year = 1991,
        month = sep,
       volume = {135},
       number = {1},
        pages = {99-105},
          doi = {10.1007/BF00146701},
       adsurl = {https://ui.adsabs.harvard.edu/abs/1991SoPh..135...99B}
}

@ARTICLE{Ding_2003,
       author = {{Ding}, M.~D. and {Chen}, Q.~R. and {Li}, J.~P. and {Chen}, P.~F.},
        title = "{H{\ensuremath{\alpha}} and Hard X-Ray Observations of a Two-Ribbon Flare Associated with a Filament Eruption}",
      journal = {\apj},
         year = 2003,
        month = nov,
       volume = {598},
       number = {1},
        pages = {683-688},
          doi = {10.1086/378877},
archivePrefix = {arXiv},
       eprint = {astro-ph/0308085},
 primaryClass = {astro-ph},
       adsurl = {https://ui.adsabs.harvard.edu/abs/2003ApJ...598..683D}
}

@ARTICLE{Dewilde_2025,
       author = {{De Wilde}, M. and {Pietrow}, A.~G.~M. and {Druett}, M.~K. and {Pastor Yabar}, A. and {Koza}, J. and {Kontogiannis}, I. and {Andriienko}, O. and {Berlicki}, A. and {Brunvoll}, A.~R. and {de la Cruz Rodr{\'\i}guez}, J. and {Faber}, J.~T. and {Joshi}, R. and {Kuridze}, D. and {N{\'o}brega-Siverio}, D. and {Rouppe van der Voort}, L.~H.~M. and {Ryb{\'a}k}, J. and {Scullion}, E. and {Silva}, A.~M. and {Vashalomidze}, Z. and {Ar{\'e}valo}, A. Vicente and {Wi{\'s}niewska}, A. and {Yadav}, R. and {Zaqarashvili}, T.~V. and {Zbinden}, J. and {{\O}yre}, E.~S.},
        title = "{Synthesizing Sun-as-a-star flare spectra from high-resolution solar observations}",
      journal = {arXiv e-prints},
         year = 2025,
        month = jul,
          eid = {arXiv:2507.07967},
        pages = {arXiv:2507.07967},
          doi = {10.48550/arXiv.2507.07967},
archivePrefix = {arXiv},
       eprint = {2507.07967},
 primaryClass = {astro-ph.SR},
       adsurl = {https://ui.adsabs.harvard.edu/abs/2025arXiv250707967D}
}

@ARTICLE{Milligan_2021,
       author = {{Milligan}, Ryan O.},
        title = "{Solar Irradiance Variability Due to Solar Flares Observed in Lyman-Alpha Emission}",
      journal = {\solphys},
         year = 2021,
        month = mar,
       volume = {296},
       number = {3},
          eid = {51},
        pages = {51},
          doi = {10.1007/s11207-021-01796-3},
archivePrefix = {arXiv},
       eprint = {2102.00974},
 primaryClass = {astro-ph.SR},
       adsurl = {https://ui.adsabs.harvard.edu/abs/2021SoPh..296...51M}
}

@ARTICLE{Yashiro_2006,
       author = {{Yashiro}, S. and {Akiyama}, S. and {Gopalswamy}, N. and {Howard}, R.~A.},
        title = "{Different Power-Law Indices in the Frequency Distributions of Flares with and without Coronal Mass Ejections}",
      journal = {\apjl},
         year = 2006,
        month = oct,
       volume = {650},
       number = {2},
        pages = {L143-L146},
          doi = {10.1086/508876},
archivePrefix = {arXiv},
       eprint = {astro-ph/0609197},
 primaryClass = {astro-ph},
       adsurl = {https://ui.adsabs.harvard.edu/abs/2006ApJ...650L.143Y}
}

@ARTICLE{Wang_2007,
       author = {{Wang}, Yuming and {Zhang}, Jie},
        title = "{A Comparative Study between Eruptive X-Class Flares Associated with Coronal Mass Ejections and Confined X-Class Flares}",
      journal = {\apj},
         year = 2007,
        month = aug,
       volume = {665},
       number = {2},
        pages = {1428-1438},
          doi = {10.1086/519765},
archivePrefix = {arXiv},
       eprint = {0808.2976},
 primaryClass = {astro-ph},
       adsurl = {https://ui.adsabs.harvard.edu/abs/2007ApJ...665.1428W}
}

@ARTICLE{Zhang_2022,
       author = {{Zhang}, Yanjie and {Zhang}, Qingmin and {Song}, Dechao and {Li}, Shuting and {Dai}, Jun and {Xu}, Zhe and {Ji}, Haisheng},
        title = "{Statistical Analysis of Circular-ribbon Flares}",
      journal = {The Astrophysical Journal Supplement Series},
         year = 2022,
        month = may,
       volume = {260},
       number = {1},
          eid = {19},
        pages = {19},
          doi = {10.3847/1538-4365/ac5f4c},
archivePrefix = {arXiv},
       eprint = {2203.12819},
 primaryClass = {astro-ph.SR},
       adsurl = {https://ui.adsabs.harvard.edu/abs/2022ApJS..260...19Z}
}

@ARTICLE{Inoue_2013,
       author = {{Inoue}, S. and {Hayashi}, K. and {Shiota}, D. and {Magara}, T. and {Choe}, G.~S.},
        title = "{Magnetic Structure Producing X- and M-class Solar Flares in Solar Active Region 11158}",
      journal = {\apj},
         year = 2013,
        month = jun,
       volume = {770},
       number = {1},
          eid = {79},
        pages = {79},
          doi = {10.1088/0004-637X/770/1/79},
archivePrefix = {arXiv},
       eprint = {1304.8073},
 primaryClass = {astro-ph.SR},
       adsurl = {https://ui.adsabs.harvard.edu/abs/2013ApJ...770...79I}
}

@ARTICLE{Li_2024,
       author = {{Li}, Ting and {Zheng}, Yanfang and {Li}, Xuefeng and {Hou}, Yijun and {Li}, Xuebao and {Zhang}, Yining and {Chen}, Anqin},
        title = "{Survey of Magnetic Field Parameters Associated with Large Solar Flares}",
      journal = {\apj},
         year = 2024,
        month = apr,
       volume = {964},
       number = {2},
          eid = {159},
        pages = {159},
          doi = {10.3847/1538-4357/ad2e90},
archivePrefix = {arXiv},
       eprint = {2402.18890},
 primaryClass = {astro-ph.SR},
       adsurl = {https://ui.adsabs.harvard.edu/abs/2024ApJ...964..159L}
}

@ARTICLE{Milligan_2020,
       author = {{Milligan}, Ryan O. and {Hudson}, Hugh S. and {Chamberlin}, Phillip C. and {Hannah}, Iain G. and {Hayes}, Laura A.},
        title = "{Lyman-alpha Variability During Solar Flares Over Solar Cycle 24 Using GOES-15/EUVS-E}",
      journal = {Space Weather},
         year = 2020,
        month = jul,
       volume = {18},
       number = {7},
          eid = {e02331},
        pages = {e02331},
          doi = {10.1029/2019SW002331},
archivePrefix = {arXiv},
       eprint = {1910.01364},
 primaryClass = {astro-ph.SR},
       adsurl = {https://ui.adsabs.harvard.edu/abs/2020SpWea..1802331M}
}

@ARTICLE{Greatorex_2024,
       author = {{Greatorex}, Harry J. and {Milligan}, Ryan O. and {Dammasch}, Ingolf E.},
        title = "{On the Instrumental Discrepancies in Lyman-Alpha Observations of Solar Flares}",
      journal = {\solphys},
         year = 2024,
        month = nov,
       volume = {299},
       number = {11},
          eid = {162},
        pages = {162},
          doi = {10.1007/s11207-024-02407-7},
archivePrefix = {arXiv},
       eprint = {2411.00736},
 primaryClass = {astro-ph.SR},
       adsurl = {https://ui.adsabs.harvard.edu/abs/2024SoPh..299..162G}
}

@ARTICLE{Mccauley_2015,
       author = {{McCauley}, P.~I. and {Su}, Y.~N. and {Schanche}, N. and {Evans}, K.~E. and {Su}, C. and {McKillop}, S. and {Reeves}, K.~K.},
        title = "{Prominence and Filament Eruptions Observed by the Solar Dynamics Observatory: Statistical Properties, Kinematics, and Online Catalog}",
      journal = {\solphys},
         year = 2015,
        month = jun,
       volume = {290},
       number = {6},
        pages = {1703-1740},
          doi = {10.1007/s11207-015-0699-7},
archivePrefix = {arXiv},
       eprint = {1505.02090},
 primaryClass = {astro-ph.SR},
       adsurl = {https://ui.adsabs.harvard.edu/abs/2015SoPh..290.1703M}
}

@ARTICLE{Culhane_2007,
       author = {{Culhane}, J.~L. and {Harra}, L.~K. and {James}, A.~M. and {Al-Janabi}, K. and {Bradley}, L.~J. and {Chaudry}, R.~A. and {Rees}, K. and {Tandy}, J.~A. and {Thomas}, P. and {Whillock}, M.~C.~R. and {Winter}, B. and {Doschek}, G.~A. and {Korendyke}, C.~M. and {Brown}, C.~M. and {Myers}, S. and {Mariska}, J. and {Seely}, J. and {Lang}, J. and {Kent}, B.~J. and {Shaughnessy}, B.~M. and {Young}, P.~R. and {Simnett}, G.~M. and {Castelli}, C.~M. and {Mahmoud}, S. and {Mapson-Menard}, H. and {Probyn}, B.~J. and {Thomas}, R.~J. and {Davila}, J. and {Dere}, K. and {Windt}, D. and {Shea}, J. and {Hagood}, R. and {Moye}, R. and {Hara}, H. and {Watanabe}, T. and {Matsuzaki}, K. and {Kosugi}, T. and {Hansteen}, V. and {Wikstol}, {\O}.},
        title = "{The EUV Imaging Spectrometer for Hinode}",
      journal = {\solphys},
         year = 2007,
        month = jun,
       volume = {243},
       number = {1},
        pages = {19-61},
          doi = {10.1007/s01007-007-0293-1},
       adsurl = {https://ui.adsabs.harvard.edu/abs/2007SoPh..243...19C}
}

@ARTICLE{Depontieu_2014,
       author = {{De Pontieu}, B. and {Title}, A.~M. and {Lemen}, J.~R. and {Kushner}, G.~D. and {Akin}, D.~J. and {Allard}, B. and {Berger}, T. and {Boerner}, P. and {Cheung}, M. and {Chou}, C. and {Drake}, J.~F. and {Duncan}, D.~W. and {Freeland}, S. and {Heyman}, G.~F. and {Hoffman}, C. and {Hurlburt}, N.~E. and {Lindgren}, R.~W. and {Mathur}, D. and {Rehse}, R. and {Sabolish}, D. and {Seguin}, R. and {Schrijver}, C.~J. and {Tarbell}, T.~D. and {W{\"u}lser}, J. -P. and {Wolfson}, C.~J. and {Yanari}, C. and {Mudge}, J. and {Nguyen-Phuc}, N. and {Timmons}, R. and {van Bezooijen}, R. and {Weingrod}, I. and {Brookner}, R. and {Butcher}, G. and {Dougherty}, B. and {Eder}, J. and {Knagenhjelm}, V. and {Larsen}, S. and {Mansir}, D. and {Phan}, L. and {Boyle}, P. and {Cheimets}, P.~N. and {DeLuca}, E.~E. and {Golub}, L. and {Gates}, R. and {Hertz}, E. and {McKillop}, S. and {Park}, S. and {Perry}, T. and {Podgorski}, W.~A. and {Reeves}, K. and {Saar}, S. and {Testa}, P. and {Tian}, H. and {Weber}, M. and {Dunn}, C. and {Eccles}, S. and {Jaeggli}, S.~A. and {Kankelborg}, C.~C. and {Mashburn}, K. and {Pust}, N. and {Springer}, L. and {Carvalho}, R. and {Kleint}, L. and {Marmie}, J. and {Mazmanian}, E. and {Pereira}, T.~M.~D. and {Sawyer}, S. and {Strong}, J. and {Worden}, S.~P. and {Carlsson}, M. and {Hansteen}, V.~H. and {Leenaarts}, J. and {Wiesmann}, M. and {Aloise}, J. and {Chu}, K. -C. and {Bush}, R.~I. and {Scherrer}, P.~H. and {Brekke}, P. and {Martinez-Sykora}, J. and {Lites}, B.~W. and {McIntosh}, S.~W. and {Uitenbroek}, H. and {Okamoto}, T.~J. and {Gummin}, M.~A. and {Auker}, G. and {Jerram}, P. and {Pool}, P. and {Waltham}, N.},
        title = "{The Interface Region Imaging Spectrograph (IRIS)}",
      journal = {\solphys},
         year = 2014,
        month = jul,
       volume = {289},
       number = {7},
        pages = {2733-2779},
          doi = {10.1007/s11207-014-0485-y},
archivePrefix = {arXiv},
       eprint = {1401.2491},
 primaryClass = {astro-ph.SR},
       adsurl = {https://ui.adsabs.harvard.edu/abs/2014SoPh..289.2733D}
}

@ARTICLE{Allred_2015,
       author = {{Allred}, Joel C. and {Kowalski}, Adam F. and {Carlsson}, Mats},
        title = "{A Unified Computational Model for Solar and Stellar Flares}",
      journal = {\apj},
         year = 2015,
        month = aug,
       volume = {809},
       number = {1},
          eid = {104},
        pages = {104},
          doi = {10.1088/0004-637X/809/1/104},
archivePrefix = {arXiv},
       eprint = {1507.04375},
 primaryClass = {astro-ph.SR},
       adsurl = {https://ui.adsabs.harvard.edu/abs/2015ApJ...809..104A}
}

@INPROCEEDINGS{Russell_2024,
       author = {{Russell}, Alexander J.~B.},
        title = "{Alfv{\'e}n Waves in Solar Flares}",
    booktitle = {Alfv{\'e}n Waves Across Heliophysics: Progress, Challenges, and Opportunities},
         year = 2024,
       editor = {{Keiling}, Andreas},
       volume = {285},
        month = jan,
        pages = {39-73},
          doi = {10.1002/9781394195985.ch3},
archivePrefix = {arXiv},
       eprint = {2311.02144},
 primaryClass = {astro-ph.SR},
       adsurl = {https://ui.adsabs.harvard.edu/abs/2024GMS...285...39R}
}

@ARTICLE{Jenkins_2023,
       author = {{Jenkins}, J.~M. and {Osborne}, C.~M.~J. and {Keppens}, R.},
        title = "{1.5D non-LTE spectral synthesis of a 3D filament and prominence simulation}",
      journal = {\aap},
         year = 2023,
        month = feb,
       volume = {670},
          eid = {A179},
        pages = {A179},
          doi = {10.1051/0004-6361/202244868},
archivePrefix = {arXiv},
       eprint = {2211.14869},
 primaryClass = {astro-ph.SR},
       adsurl = {https://ui.adsabs.harvard.edu/abs/2023A&A...670A.179J}
}

@ARTICLE{Watanabe_2021,
       author = {{Watanabe}, Kyoko and {Jin}, Hidekatsu and {Nishimoto}, Shohei and {Imada}, Shinsuke and {Kawai}, Toshiki and {Kawate}, Tomoko and {Otsuka}, Yuichi and {Shinbori}, Atsuki and {Tsugawa}, Takuya and {Nishioka}, Michi},
        title = "{Model-based reproduction and validation of the total spectra of a solar flare and their impact on the global environment at the X9.3 event of September 6, 2017}",
      journal = {Earth, Planets and Space},
         year = 2021,
        month = dec,
       volume = {73},
       number = {1},
          eid = {96},
        pages = {96},
          doi = {10.1186/s40623-021-01376-6},
       adsurl = {https://ui.adsabs.harvard.edu/abs/2021EP&S...73...96W}
}

@ARTICLE{Ohare_2025,
       author = {{O'Hare}, Aisling N. and {Bekker}, Susanna and {Greatorex}, Harry J. and {Milligan}, Ryan O.},
        title = "{Investigating a Characteristic Time Lag in the Ionospheric F-region's Response to Solar Flares}",
      journal = {arXiv e-prints},
         year = 2025,
        month = aug,
          eid = {arXiv:2508.03425},
        pages = {arXiv:2508.03425},
archivePrefix = {arXiv},
       eprint = {2508.03425},
 primaryClass = {astro-ph.SR},
       adsurl = {https://ui.adsabs.harvard.edu/abs/2025arXiv250803425O}
}

@ARTICLE{Argiroffi_2019,
       author = {{Argiroffi}, C. and {Reale}, F. and {Drake}, J.~J. and {Ciaravella}, A. and {Testa}, P. and {Bonito}, R. and {Miceli}, M. and {Orlando}, S. and {Peres}, G.},
        title = "{A stellar flare-coronal mass ejection event revealed by X-ray plasma motions}",
      journal = {Nature Astronomy},
         year = 2019,
        month = may,
       volume = {3},
        pages = {742-748},
          doi = {10.1038/s41550-019-0781-4},
archivePrefix = {arXiv},
       eprint = {1905.11325},
 primaryClass = {astro-ph.SR},
       adsurl = {https://ui.adsabs.harvard.edu/abs/2019NatAs...3..742A}
}

@ARTICLE{Hudson_2021,
       author = {{Hudson}, Hugh S. and {Sim{\~o}es}, Paulo J.~A. and {Fletcher}, Lyndsay and {Hayes}, Laura A. and {Hannah}, Iain G.},
        title = "{Hot X-ray onsets of solar flares}",
      journal = {\mnras},
         year = 2021,
        month = feb,
       volume = {501},
       number = {1},
        pages = {1273-1281},
          doi = {10.1093/mnras/staa3664},
archivePrefix = {arXiv},
       eprint = {2007.05310},
 primaryClass = {astro-ph.SR},
       adsurl = {https://ui.adsabs.harvard.edu/abs/2021MNRAS.501.1273H}
}

@ARTICLE{Battaglia_2023,
       author = {{Battaglia}, Andrea Francesco and {Hudson}, Hugh and {Warmuth}, Alexander and {Collier}, Hannah and {Jeffrey}, Natasha L.~S. and {Caspi}, Amir and {Dickson}, Ewan C.~M. and {Saqri}, Jonas and {Purkhart}, Stefan and {Veronig}, Astrid M. and {Harra}, Louise and {Krucker}, S{\"a}m},
        title = "{The existence of hot X-ray onsets in solar flares}",
      journal = {\aap},
         year = 2023,
        month = nov,
       volume = {679},
          eid = {A139},
        pages = {A139},
          doi = {10.1051/0004-6361/202347706},
archivePrefix = {arXiv},
       eprint = {2310.04234},
 primaryClass = {astro-ph.SR},
       adsurl = {https://ui.adsabs.harvard.edu/abs/2023A&A...679A.139B}
}

@ARTICLE{Hayes_2024,
       author = {{Hayes}, L.~A. and {Krucker}, S. and {Collier}, H. and {Ryan}, D.},
        title = "{High-energy insights from an escaping coronal mass ejection with Solar Orbiter/STIX observations}",
      journal = {\aap},
         year = 2024,
        month = nov,
       volume = {691},
          eid = {A190},
        pages = {A190},
          doi = {10.1051/0004-6361/202450882},
archivePrefix = {arXiv},
       eprint = {2408.14194},
 primaryClass = {astro-ph.SR},
       adsurl = {https://ui.adsabs.harvard.edu/abs/2024A&A...691A.190H}
}

@ARTICLE{Emslie_2012,
       author = {{Emslie}, A.~G. and {Dennis}, B.~R. and {Shih}, A.~Y. and {Chamberlin}, P.~C. and {Mewaldt}, R.~A. and {Moore}, C.~S. and {Share}, G.~H. and {Vourlidas}, A. and {Welsch}, B.~T.},
        title = "{Global Energetics of Thirty-eight Large Solar Eruptive Events}",
      journal = {\apj},
         year = 2012,
        month = nov,
       volume = {759},
       number = {1},
          eid = {71},
        pages = {71},
          doi = {10.1088/0004-637X/759/1/71},
archivePrefix = {arXiv},
       eprint = {1209.2654},
 primaryClass = {astro-ph.SR},
       adsurl = {https://ui.adsabs.harvard.edu/abs/2012ApJ...759...71E}
}

@ARTICLE{Mierla_2022,
       author = {{Mierla}, M. and {Zhukov}, A.~N. and {Berghmans}, D. and {Parenti}, S. and {Auch{\`e}re}, F. and {Heinzel}, P. and {Seaton}, D.~B. and {Palmerio}, E. and {Jej{\v{c}}i{\v{c}}}, S. and {Janssens}, J. and {Kraaikamp}, E. and {Nicula}, B. and {Long}, D.~M. and {Hayes}, L.~A. and {Jebaraj}, I.~C. and {Talpeanu}, D. -C. and {D'Huys}, E. and {Dolla}, L. and {Gissot}, S. and {Magdaleni{\'c}}, J. and {Rodriguez}, L. and {Shestov}, S. and {Stegen}, K. and {Verbeeck}, C. and {Sasso}, C. and {Romoli}, M. and {Andretta}, V.},
        title = "{Prominence eruption observed in He II 304 {\r{A}} up to >6 R$_{{\ensuremath{\odot}}}$ by EUI/FSI aboard Solar Orbiter}",
      journal = {\aap},
         year = 2022,
        month = jun,
       volume = {662},
          eid = {L5},
        pages = {L5},
          doi = {10.1051/0004-6361/202244020},
archivePrefix = {arXiv},
       eprint = {2205.15214},
 primaryClass = {astro-ph.SR},
       adsurl = {https://ui.adsabs.harvard.edu/abs/2022A&A...662L...5M}
}

@ARTICLE{Lu_2021,
       author = {{Lu}, Lei and {Feng}, Li and {Li}, Dong and {Ying}, Beili and {Li}, Hui and {Gan}, Weiqun and {Li}, Youping and {Zhou}, Jiujiu},
        title = "{Catalog and Statistical Examinations of Ly{\ensuremath{\alpha}} Solar Flares from GOES/EUVS-E Measurements}",
      journal = {The Astrophysical Journal Supplement Series},
         year = 2021,
        month = mar,
       volume = {253},
       number = {1},
          eid = {29},
        pages = {29},
          doi = {10.3847/1538-4365/abd79b},
       adsurl = {https://ui.adsabs.harvard.edu/abs/2021ApJS..253...29L}
}

@ARTICLE{Greatorex_2023,
       author = {{Greatorex}, Harry J. and {Milligan}, Ryan O. and {Chamberlin}, Phillip C.},
        title = "{Observational Analysis of Ly{\ensuremath{\alpha}} Emission in Equivalent-magnitude Solar Flares}",
      journal = {\apj},
         year = 2023,
        month = sep,
       volume = {954},
       number = {2},
          eid = {120},
        pages = {120},
          doi = {10.3847/1538-4357/acea7f},
archivePrefix = {arXiv},
       eprint = {2306.16234},
 primaryClass = {astro-ph.SR},
       adsurl = {https://ui.adsabs.harvard.edu/abs/2023ApJ...954..120G}
}

@ARTICLE{Xue_2016,
       author = {{Xue}, Zhike and {Yan}, Xiaoli and {Cheng}, Xin and {Yang}, Liheng and {Su}, Yingna and {Kliem}, Bernhard and {Zhang}, Jun and {Liu}, Zhong and {Bi}, Yi and {Xiang}, Yongyuan and {Yang}, Kai and {Zhao}, Li},
        title = "{Observing the release of twist by magnetic reconnection in a solar filament eruption}",
      journal = {Nature Communications},
         year = 2016,
        month = jun,
       volume = {7},
          eid = {11837},
        pages = {11837},
          doi = {10.1038/ncomms11837},
       adsurl = {https://ui.adsabs.harvard.edu/abs/2016NatCo...711837X}
}

@ARTICLE{Reeves_2019,
       author = {{Reeves}, Katharine K. and {T{\"o}r{\"o}k}, Tibor and {Miki{\'c}}, Zoran and {Linker}, Jon and {Murphy}, Nicholas A.},
        title = "{Exploring Plasma Heating in the Current Sheet Region in a Three-dimensional Coronal Mass Ejection Simulation}",
      journal = {\apj},
         year = 2019,
        month = dec,
       volume = {887},
       number = {1},
          eid = {103},
        pages = {103},
          doi = {10.3847/1538-4357/ab4ce8},
archivePrefix = {arXiv},
       eprint = {1910.05386},
 primaryClass = {astro-ph.SR},
       adsurl = {https://ui.adsabs.harvard.edu/abs/2019ApJ...887..103R}
}

@ARTICLE{Veronig_2021,
       author = {{Veronig}, Astrid M. and {Odert}, Petra and {Leitzinger}, Martin and {Dissauer}, Karin and {Fleck}, Nikolaus C. and {Hudson}, Hugh S.},
        title = "{Indications of stellar coronal mass ejections through coronal dimmings}",
      journal = {Nature Astronomy},
         year = 2021,
        month = jan,
       volume = {5},
        pages = {697-706},
          doi = {10.1038/s41550-021-01345-9},
archivePrefix = {arXiv},
       eprint = {2110.12029},
 primaryClass = {astro-ph.SR},
       adsurl = {https://ui.adsabs.harvard.edu/abs/2021NatAs...5..697V}
}

@ARTICLE{Gunn_1994,
       author = {{Gunn}, A.~G. and {Doyle}, J.~G. and {Mathioudakis}, M. and {Houdebine}, E.~R. and {Avgoloupis}, S.},
        title = "{High-velocity evaporation during a flare on AT Microscopii}",
      journal = {\aap},
         year = 1994,
        month = may,
       volume = {285},
        pages = {489-496},
       adsurl = {https://ui.adsabs.harvard.edu/abs/1994A&A...285..489G}
}

@ARTICLE{Berdyugina_1999,
       author = {{Berdyugina}, S.~V. and {Ilyin}, I. and {Tuominen}, I.},
        title = "{The active RS Canum Venaticorum binary II Pegasi. III. Chromospheric emission and flares in 1994-1996}",
      journal = {\aap},
         year = 1999,
        month = sep,
       volume = {349},
        pages = {863-872},
       adsurl = {https://ui.adsabs.harvard.edu/abs/1999A&A...349..863B}
}

@ARTICLE{Wang_2024,
       author = {{Wang}, J. and {Mao}, X. and {Gao}, C. and {Liu}, H.~Y. and {Li}, H.~L. and {Pan}, H.~W. and {Wu}, C. and {Liu}, Y. and {Li}, G.~W. and {Xin}, L.~P. and {Jin}, S. and {Xu}, D.~W. and {Liang}, E.~W. and {Yuan}, W.~M. and {Wei}, J.~Y.},
        title = "{Potential Chromospheric Evaporation in the M Dwarf's Flare Triggered by Einstein Probe Mission}",
      journal = {\aj},
         year = 2024,
        month = dec,
       volume = {168},
       number = {6},
          eid = {261},
        pages = {261},
          doi = {10.3847/1538-3881/ad83b4},
archivePrefix = {arXiv},
       eprint = {2410.03114},
 primaryClass = {astro-ph.SR},
       adsurl = {https://ui.adsabs.harvard.edu/abs/2024AJ....168..261W}
}

@ARTICLE{Varady_2010,
       author = {{Varady}, Michal and {Kasparova}, Jana and {Moravec}, Zden{\v{e}}k and {Heinzel}, Petr and {Karlicky}, Marian},
        title = "{Modeling of Solar Flare Plasma and Its Radiation}",
      journal = {IEEE Transactions on Plasma Science},
         year = 2010,
        month = sep,
       volume = {38},
       number = {9},
        pages = {2249-2253},
          doi = {10.1109/TPS.2010.2057449},
       adsurl = {https://ui.adsabs.harvard.edu/abs/2010ITPS...38.2249V}
}

@INPROCEEDINGS{Heinzel_2016,
       author = {{Heinzel}, Petr and {Ka{\v{s}}parov{\'a}}, Jana and {Varady}, Michal and {Karlick{\'y}}, Marian and {Moravec}, Zden{\v{e}}k},
        title = "{Numerical RHD simulations of flaring chromosphere with Flarix}",
    booktitle = {Solar and Stellar Flares and their Effects on Planets},
         year = 2016,
       editor = {{Kosovichev}, A.~G. and {Hawley}, S.~L. and {Heinzel}, P.},
       series = {IAU Symposium},
       volume = {320},
        month = jan,
        pages = {233-238},
          doi = {10.1017/S1743921316000363},
archivePrefix = {arXiv},
       eprint = {1602.00016},
 primaryClass = {astro-ph.SR},
       adsurl = {https://ui.adsabs.harvard.edu/abs/2016IAUS..320..233H}
}

@ARTICLE{Druett_2018,
       author = {{Druett}, M.~K. and {Zharkova}, V.~V.},
        title = "{HYDRO2GEN: Non-thermal hydrogen Balmer and Paschen emission in solar flares generated by electron beams}",
      journal = {\aap},
         year = 2018,
        month = mar,
       volume = {610},
          eid = {A68},
        pages = {A68},
          doi = {10.1051/0004-6361/201731053},
       adsurl = {https://ui.adsabs.harvard.edu/abs/2018A&A...610A..68D}
}

@ARTICLE{Druett_2019,
       author = {{Druett}, M.~K. and {Zharkova}, V.~V.},
        title = "{Non-thermal hydrogen Lyman line and continuum emission in solar flares generated by electron beams}",
      journal = {\aap},
         year = 2019,
        month = mar,
       volume = {623},
          eid = {A20},
        pages = {A20},
          doi = {10.1051/0004-6361/201732427},
       adsurl = {https://ui.adsabs.harvard.edu/abs/2019A&A...623A..20D}
}

@ARTICLE{Allred_2005,
       author = {{Allred}, Joel C. and {Hawley}, Suzanne L. and {Abbett}, William P. and {Carlsson}, Mats},
        title = "{Radiative Hydrodynamic Models of the Optical and Ultraviolet Emission from Solar Flares}",
      journal = {\apj},
         year = 2005,
        month = sep,
       volume = {630},
       number = {1},
        pages = {573-586},
          doi = {10.1086/431751},
archivePrefix = {arXiv},
       eprint = {astro-ph/0507335},
 primaryClass = {astro-ph},
       adsurl = {https://ui.adsabs.harvard.edu/abs/2005ApJ...630..573A}
}

@ARTICLE{Druett_2024,
       author = {{Druett}, Malcolm and {Ruan}, Wenzhi and {Keppens}, Rony},
        title = "{Exploring self-consistent 2.5D flare simulations with MPI-AMRVAC}",
      journal = {\aap},
         year = 2024,
        month = apr,
       volume = {684},
          eid = {A171},
        pages = {A171},
          doi = {10.1051/0004-6361/202347600},
archivePrefix = {arXiv},
       eprint = {2310.09939},
 primaryClass = {astro-ph.SR},
       adsurl = {https://ui.adsabs.harvard.edu/abs/2024A&A...684A.171D}
}

@ARTICLE{Jenkins_2021,
       author = {{Jenkins}, J.~M. and {Keppens}, R.},
        title = "{Prominence formation by levitation-condensation at extreme resolutions}",
      journal = {\aap},
         year = 2021,
        month = feb,
       volume = {646},
          eid = {A134},
        pages = {A134},
          doi = {10.1051/0004-6361/202039630},
archivePrefix = {arXiv},
       eprint = {2011.13428},
 primaryClass = {astro-ph.SR},
       adsurl = {https://ui.adsabs.harvard.edu/abs/2021A&A...646A.134J}
}

@ARTICLE{Jenkins_2022,
       author = {{Jenkins}, Jack M. and {Keppens}, Rony},
        title = "{Resolving the solar prominence/filament paradox using the magnetic Rayleigh-Taylor instability}",
      journal = {Nature Astronomy},
         year = 2022,
        month = jul,
       volume = {6},
        pages = {942-950},
          doi = {10.1038/s41550-022-01705-z},
       adsurl = {https://ui.adsabs.harvard.edu/abs/2022NatAs...6..942J}
}

@ARTICLE{Lemen_2012,
       author = {{Lemen}, James R. and {Title}, Alan M. and {Akin}, David J. and {Boerner}, Paul F. and {Chou}, Catherine and {Drake}, Jerry F. and {Duncan}, Dexter W. and {Edwards}, Christopher G. and {Friedlaender}, Frank M. and {Heyman}, Gary F. and {Hurlburt}, Neal E. and {Katz}, Noah L. and {Kushner}, Gary D. and {Levay}, Michael and {Lindgren}, Russell W. and {Mathur}, Dnyanesh P. and {McFeaters}, Edward L. and {Mitchell}, Sarah and {Rehse}, Roger A. and {Schrijver}, Carolus J. and {Springer}, Larry A. and {Stern}, Robert A. and {Tarbell}, Theodore D. and {Wuelser}, Jean-Pierre and {Wolfson}, C. Jacob and {Yanari}, Carl and {Bookbinder}, Jay A. and {Cheimets}, Peter N. and {Caldwell}, David and {Deluca}, Edward E. and {Gates}, Richard and {Golub}, Leon and {Park}, Sang and {Podgorski}, William A. and {Bush}, Rock I. and {Scherrer}, Philip H. and {Gummin}, Mark A. and {Smith}, Peter and {Auker}, Gary and {Jerram}, Paul and {Pool}, Peter and {Soufli}, Regina and {Windt}, David L. and {Beardsley}, Sarah and {Clapp}, Matthew and {Lang}, James and {Waltham}, Nicholas},
        title = "{The Atmospheric Imaging Assembly (AIA) on the Solar Dynamics Observatory (SDO)}",
      journal = {\solphys},
         year = 2012,
        month = jan,
       volume = {275},
       number = {1-2},
        pages = {17-40},
          doi = {10.1007/s11207-011-9776-8},
       adsurl = {https://ui.adsabs.harvard.edu/abs/2012SoPh..275...17L}
}

@ARTICLE{Pesnell_2012,
       author = {{Pesnell}, W. Dean and {Thompson}, B.~J. and {Chamberlin}, P.~C.},
        title = "{The Solar Dynamics Observatory (SDO)}",
      journal = {\solphys},
         year = 2012,
        month = jan,
       volume = {275},
       number = {1-2},
        pages = {3-15},
          doi = {10.1007/s11207-011-9841-3},
       adsurl = {https://ui.adsabs.harvard.edu/abs/2012SoPh..275....3P}
}

@ARTICLE{Dominique_2013,
       author = {{Dominique}, M. and {Hochedez}, J. -F. and {Schmutz}, W. and {Dammasch}, I.~E. and {Shapiro}, A.~I. and {Kretzschmar}, M. and {Zhukov}, A.~N. and {Gillotay}, D. and {Stockman}, Y. and {BenMoussa}, A.},
        title = "{The LYRA Instrument Onboard PROBA2: Description and In-Flight Performance}",
      journal = {\solphys},
         year = 2013,
        month = aug,
       volume = {286},
       number = {1},
        pages = {21-42},
          doi = {10.1007/s11207-013-0252-5},
archivePrefix = {arXiv},
       eprint = {1302.6525},
 primaryClass = {astro-ph.IM},
       adsurl = {https://ui.adsabs.harvard.edu/abs/2013SoPh..286...21D}
}

@ARTICLE{Hochedez_2006,
       author = {{Hochedez}, J. -F. and {Schmutz}, W. and {Stockman}, Y. and {Sch{\"u}hle}, U. and {Benmoussa}, A. and {Koller}, S. and {Haenen}, K. and {Berghmans}, D. and {Defise}, J. -M. and {Halain}, J. -P. and {Theissen}, A. and {Delouille}, V. and {Slemzin}, V. and {Gillotay}, D. and {Fussen}, D. and {Dominique}, M. and {Vanhellemont}, F. and {McMullin}, D. and {Kretzschmar}, M. and {Mitrofanov}, A. and {Nicula}, B. and {Wauters}, L. and {Roth}, H. and {Rozanov}, E. and {R{\"u}edi}, I. and {Wehrli}, C. and {Soltani}, A. and {Amano}, H. and {van der Linden}, R. and {Zhukov}, A. and {Clette}, F. and {Koizumi}, S. and {Mortet}, V. and {Remes}, Z. and {Petersen}, R. and {Nesl{\'a}dek}, M. and {D'Olieslaeger}, M. and {Roggen}, J. and {Rochus}, P.},
        title = "{LYRA, a solar UV radiometer on Proba2}",
      journal = {Advances in Space Research},
         year = 2006,
        month = jan,
       volume = {37},
       number = {2},
        pages = {303-312},
          doi = {10.1016/j.asr.2005.10.041},
       adsurl = {https://ui.adsabs.harvard.edu/abs/2006AdSpR..37..303H}
}

@INPROCEEDINGS{Chamberlin_2009,
       author = {{Chamberlin}, Phillip C. and {Woods}, Thomas N. and {Eparvier}, Francis G. and {Jones}, Andrew R.},
        title = "{Next generation x-ray sensor (XRS) for the NOAA GOES-R satellite series}",
    booktitle = {Solar Physics and Space Weather Instrumentation III},
         year = 2009,
       editor = {{Fineschi}, Silvano and {Fennelly}, Judy A.},
       series = {Society of Photo-Optical Instrumentation Engineers (SPIE) Conference Series},
       volume = {7438},
        month = aug,
          eid = {743802},
        pages = {743802},
          doi = {10.1117/12.826807},
       adsurl = {https://ui.adsabs.harvard.edu/abs/2009SPIE.7438E..02C}
}

@INPROCEEDINGS{Eparvier_2009,
       author = {{Eparvier}, Francis G. and {Crotser}, David and {Jones}, Andrew R. and {McClintock}, William E. and {Snow}, Martin and {Woods}, Thomas N.},
        title = "{The Extreme Ultraviolet Sensor (EUVS) for GOES-R}",
    booktitle = {Solar Physics and Space Weather Instrumentation III},
         year = 2009,
       editor = {{Fineschi}, Silvano and {Fennelly}, Judy A.},
       series = {Society of Photo-Optical Instrumentation Engineers (SPIE) Conference Series},
       volume = {7438},
        month = aug,
          eid = {743804},
        pages = {743804},
          doi = {10.1117/12.826445},
       adsurl = {https://ui.adsabs.harvard.edu/abs/2009SPIE.7438E..04E}
}

@ARTICLE{Muller_2020,
       author = {{M{\"u}ller}, D. and {St. Cyr}, O.~C. and {Zouganelis}, I. and {Gilbert}, H.~R. and {Marsden}, R. and {Nieves-Chinchilla}, T. and {Antonucci}, E. and {Auch{\`e}re}, F. and {Berghmans}, D. and {Horbury}, T.~S. and {Howard}, R.~A. and {Krucker}, S. and {Maksimovic}, M. and {Owen}, C.~J. and {Rochus}, P. and {Rodriguez-Pacheco}, J. and {Romoli}, M. and {Solanki}, S.~K. and {Bruno}, R. and {Carlsson}, M. and {Fludra}, A. and {Harra}, L. and {Hassler}, D.~M. and {Livi}, S. and {Louarn}, P. and {Peter}, H. and {Sch{\"u}hle}, U. and {Teriaca}, L. and {del Toro Iniesta}, J.~C. and {Wimmer-Schweingruber}, R.~F. and {Marsch}, E. and {Velli}, M. and {De Groof}, A. and {Walsh}, A. and {Williams}, D.},
        title = "{The Solar Orbiter mission. Science overview}",
      journal = {\aap},
         year = 2020,
        month = oct,
       volume = {642},
          eid = {A1},
        pages = {A1},
          doi = {10.1051/0004-6361/202038467},
archivePrefix = {arXiv},
       eprint = {2009.00861},
 primaryClass = {astro-ph.SR},
       adsurl = {https://ui.adsabs.harvard.edu/abs/2020A&A...642A...1M}
}

@ARTICLE{Rochus_2020,
       author = {{Rochus}, P. and {Auch{\`e}re}, F. and {Berghmans}, D. and {Harra}, L. and {Schmutz}, W. and {Sch{\"u}hle}, U. and {Addison}, P. and {Appourchaux}, T. and {Aznar Cuadrado}, R. and {Baker}, D. and {Barbay}, J. and {Bates}, D. and {BenMoussa}, A. and {Bergmann}, M. and {Beurthe}, C. and {Borgo}, B. and {Bonte}, K. and {Bouzit}, M. and {Bradley}, L. and {B{\"u}chel}, V. and {Buchlin}, E. and {B{\"u}chner}, J. and {Cab{\'e}}, F. and {Cadiergues}, L. and {Chaigneau}, M. and {Chares}, B. and {Choque Cortez}, C. and {Coker}, P. and {Condamin}, M. and {Coumar}, S. and {Curdt}, W. and {Cutler}, J. and {Davies}, D. and {Davison}, G. and {Defise}, J. -M. and {Del Zanna}, G. and {Delmotte}, F. and {Delouille}, V. and {Dolla}, L. and {Dumesnil}, C. and {D{\"u}rig}, F. and {Enge}, R. and {Fran{\c{c}}ois}, S. and {Fourmond}, J. -J. and {Gillis}, J. -M. and {Giordanengo}, B. and {Gissot}, S. and {Green}, L.~M. and {Guerreiro}, N. and {Guilbaud}, A. and {Gyo}, M. and {Haberreiter}, M. and {Hafiz}, A. and {Hailey}, M. and {Halain}, J. -P. and {Hansotte}, J. and {Hecquet}, C. and {Heerlein}, K. and {Hellin}, M. -L. and {Hemsley}, S. and {Hermans}, A. and {Hervier}, V. and {Hochedez}, J. -F. and {Houbrechts}, Y. and {Ihsan}, K. and {Jacques}, L. and {J{\'e}r{\^o}me}, A. and {Jones}, J. and {Kahle}, M. and {Kennedy}, T. and {Klaproth}, M. and {Kolleck}, M. and {Koller}, S. and {Kotsialos}, E. and {Kraaikamp}, E. and {Langer}, P. and {Lawrenson}, A. and {Le Clech'}, J. -C. and {Lenaerts}, C. and {Liebecq}, S. and {Linder}, D. and {Long}, D.~M. and {Mampaey}, B. and {Markiewicz-Innes}, D. and {Marquet}, B. and {Marsch}, E. and {Matthews}, S. and {Mazy}, E. and {Mazzoli}, A. and {Meining}, S. and {Meltchakov}, E. and {Mercier}, R. and {Meyer}, S. and {Monecke}, M. and {Monfort}, F. and {Morinaud}, G. and {Moron}, F. and {Mountney}, L. and {M{\"u}ller}, R. and {Nicula}, B. and {Parenti}, S. and {Peter}, H. and {Pfiffner}, D. and {Philippon}, A. and {Phillips}, I. and {Plesseria}, J. -Y. and {Pylyser}, E. and {Rabecki}, F. and {Ravet-Krill}, M. -F. and {Rebellato}, J. and {Renotte}, E. and {Rodriguez}, L. and {Roose}, S. and {Rosin}, J. and {Rossi}, L. and {Roth}, P. and {Rouesnel}, F. and {Roulliay}, M. and {Rousseau}, A. and {Ruane}, K. and {Scanlan}, J. and {Schlatter}, P. and {Seaton}, D.~B. and {Silliman}, K. and {Smit}, S. and {Smith}, P.~J. and {Solanki}, S.~K. and {Spescha}, M. and {Spencer}, A. and {Stegen}, K. and {Stockman}, Y. and {Szwec}, N. and {Tamiatto}, C. and {Tandy}, J. and {Teriaca}, L. and {Theobald}, C. and {Tychon}, I. and {van Driel-Gesztelyi}, L. and {Verbeeck}, C. and {Vial}, J. -C. and {Werner}, S. and {West}, M.~J. and {Westwood}, D. and {Wiegelmann}, T. and {Willis}, G. and {Winter}, B. and {Zerr}, A. and {Zhang}, X. and {Zhukov}, A.~N.},
        title = "{The Solar Orbiter EUI instrument: The Extreme Ultraviolet Imager}",
      journal = {\aap},
         year = 2020,
        month = oct,
       volume = {642},
          eid = {A8},
        pages = {A8},
          doi = {10.1051/0004-6361/201936663},
       adsurl = {https://ui.adsabs.harvard.edu/abs/2020A&A...642A...8R}
}

@ARTICLE{Zhang_2019,
       author = {{Zhang}, Zhe and {Chen}, Deng-Yi and {Wu}, Jian and {Chang}, Jin and {Hu}, Yi-Ming and {Su}, Yang and {Zhang}, Yan and {Wang}, Jian-Ping and {Liang}, Yao-Ming and {Ma}, Tao and {Guo}, Jian-Hua and {Cai}, Ming-Sheng and {Zhang}, Yong-Qiang and {Huang}, Yong-Yi and {Peng}, Xiao-Yan and {Tang}, Zong-Bin and {Zhao}, Xuan and {Zhou}, Hong-He and {Wang}, Lian-Guo and {Song}, Jing-Xing and {Ma}, Miao and {Xu}, Guang-Zhou and {Yang}, Jian-Feng and {Lu}, Di and {He}, Ying-Hong and {Tao}, Jin-You and {Ma}, Xiao-Long and {Lv}, Bao-Gang and {Bai}, Yan-Ping and {Cao}, Cai-Xia and {Huang}, Yu and {Gan}, Wei-Qun},
        title = "{Hard X-ray Imager (HXI) onboard the ASO-S mission}",
      journal = {Research in Astronomy and Astrophysics},
         year = 2019,
        month = nov,
       volume = {19},
       number = {11},
          eid = {160},
        pages = {160},
          doi = {10.1088/1674-4527/19/11/160},
       adsurl = {https://ui.adsabs.harvard.edu/abs/2019RAA....19..160Z}
}

@ARTICLE{Gan_2019,
       author = {{Gan}, Wei-Qun and {Zhu}, Cheng and {Deng}, Yuan-Yong and {Li}, Hui and {Su}, Yang and {Zhang}, Hai-Ying and {Chen}, Bo and {Zhang}, Zhe and {Wu}, Jian and {Deng}, Lei and {Huang}, Yu and {Yang}, Jian-Feng and {Cui}, Ji-Jun and {Chang}, Jin and {Wang}, Chi and {Wu}, Ji and {Yin}, Zeng-Shan and {Chen}, Wen and {Fang}, Cheng and {Yan}, Yi-Hua and {Lin}, Jun and {Xiong}, Wei-Ming and {Chen}, Bin and {Bao}, Hai-Chao and {Cao}, Cai-Xia and {Bai}, Yan-Ping and {Wang}, Tao and {Chen}, Bing-Long and {Li}, Xin-Yu and {Zhang}, Ye and {Feng}, Li and {Su}, Jiang-Tao and {Li}, Ying and {Chen}, Wei and {Li}, You-Ping and {Su}, Ying-Na and {Wu}, Hai-Yan and {Gu}, Mei and {Huang}, Lei and {Tang}, Xue-Jun},
        title = "{Advanced Space-based Solar Observatory (ASO-S): an overview}",
      journal = {Research in Astronomy and Astrophysics},
         year = 2019,
        month = nov,
       volume = {19},
       number = {11},
          eid = {156},
        pages = {156},
          doi = {10.1088/1674-4527/19/11/156},
       adsurl = {https://ui.adsabs.harvard.edu/abs/2019RAA....19..156G}
}

@ARTICLE{Howard_2008,
       author = {{Howard}, R.~A. and {Moses}, J.~D. and {Vourlidas}, A. and {Newmark}, J.~S. and {Socker}, D.~G. and {Plunkett}, S.~P. and {Korendyke}, C.~M. and {Cook}, J.~W. and {Hurley}, A. and {Davila}, J.~M. and {Thompson}, W.~T. and {St Cyr}, O.~C. and {Mentzell}, E. and {Mehalick}, K. and {Lemen}, J.~R. and {Wuelser}, J.~P. and {Duncan}, D.~W. and {Tarbell}, T.~D. and {Wolfson}, C.~J. and {Moore}, A. and {Harrison}, R.~A. and {Waltham}, N.~R. and {Lang}, J. and {Davis}, C.~J. and {Eyles}, C.~J. and {Mapson-Menard}, H. and {Simnett}, G.~M. and {Halain}, J.~P. and {Defise}, J.~M. and {Mazy}, E. and {Rochus}, P. and {Mercier}, R. and {Ravet}, M.~F. and {Delmotte}, F. and {Auchere}, F. and {Delaboudiniere}, J.~P. and {Bothmer}, V. and {Deutsch}, W. and {Wang}, D. and {Rich}, N. and {Cooper}, S. and {Stephens}, V. and {Maahs}, G. and {Baugh}, R. and {McMullin}, D. and {Carter}, T.},
        title = "{Sun Earth Connection Coronal and Heliospheric Investigation (SECCHI)}",
      journal = {\ssr},
         year = 2008,
        month = apr,
       volume = {136},
       number = {1-4},
        pages = {67-115},
          doi = {10.1007/s11214-008-9341-4},
       adsurl = {https://ui.adsabs.harvard.edu/abs/2008SSRv..136...67H}
}

@INPROCEEDINGS{Wuesler_2004,
       author = {{Wuelser}, Jean-Pierre and {Lemen}, James R. and {Tarbell}, Theodore D. and {Wolfson}, C.~J. and {Cannon}, Joseph C. and {Carpenter}, Brock A. and {Duncan}, Dexter W. and {Gradwohl}, Glenn S. and {Meyer}, Syndie B. and {Moore}, Augustus S. and {Navarro}, Rosemarie L. and {Pearson}, J.~D. and {Rossi}, George R. and {Springer}, Larry A. and {Howard}, Russell A. and {Moses}, John D. and {Newmark}, Jeffrey S. and {Delaboudiniere}, Jean-Pierre and {Artzner}, Guy E. and {Auchere}, Frederic and {Bougnet}, Marie and {Bouyries}, Philippe and {Bridou}, Francoise and {Clotaire}, Jean-Yves and {Colas}, Gerard and {Delmotte}, Franck and {Jerome}, Arnaud and {Lamare}, Michel and {Mercier}, Raymond and {Mullot}, Michel and {Ravet}, Marie-Francoise and {Song}, Xueyan and {Bothmer}, Volker and {Deutsch}, Werner},
        title = "{EUVI: the STEREO-SECCHI extreme ultraviolet imager}",
    booktitle = {Telescopes and Instrumentation for Solar Astrophysics},
         year = 2004,
       editor = {{Fineschi}, Silvano and {Gummin}, Mark A.},
       series = {Society of Photo-Optical Instrumentation Engineers (SPIE) Conference Series},
       volume = {5171},
        month = feb,
        pages = {111-122},
          doi = {10.1117/12.506877},
       adsurl = {https://ui.adsabs.harvard.edu/abs/2004SPIE.5171..111W}
}

@ARTICLE{Kaiser_2008,
       author = {{Kaiser}, M.~L. and {Kucera}, T.~A. and {Davila}, J.~M. and {St. Cyr}, O.~C. and {Guhathakurta}, M. and {Christian}, E.},
        title = "{The STEREO Mission: An Introduction}",
      journal = {\ssr},
         year = 2008,
        month = apr,
       volume = {136},
       number = {1-4},
        pages = {5-16},
          doi = {10.1007/s11214-007-9277-0},
       adsurl = {https://ui.adsabs.harvard.edu/abs/2008SSRv..136....5K}
}

@ARTICLE{Fletcher_2011,
       author = {{Fletcher}, L. and {Dennis}, B.~R. and {Hudson}, H.~S. and {Krucker}, S. and {Phillips}, K. and {Veronig}, A. and {Battaglia}, M. and {Bone}, L. and {Caspi}, A. and {Chen}, Q. and {Gallagher}, P. and {Grigis}, P.~T. and {Ji}, H. and {Liu}, W. and {Milligan}, R.~O. and {Temmer}, M.},
        title = "{An Observational Overview of Solar Flares}",
      journal = {\ssr},
         year = 2011,
        month = sep,
       volume = {159},
       number = {1-4},
        pages = {19-106},
          doi = {10.1007/s11214-010-9701-8},
archivePrefix = {arXiv},
       eprint = {1109.5932},
 primaryClass = {astro-ph.SR},
       adsurl = {https://ui.adsabs.harvard.edu/abs/2011SSRv..159...19F}
}

@ARTICLE{Cannon_2013,
       author = {{Cannon}, Paul S.},
        title = "{Extreme Space Weather{\textemdash}A Report Published by the UK Royal Academy of Engineering}",
      journal = {Space Weather},
         year = 2013,
        month = apr,
       volume = {11},
       number = {4},
        pages = {138-139},
          doi = {10.1002/swe.20032},
       adsurl = {https://ui.adsabs.harvard.edu/abs/2013SpWea..11..138C}
}

@ARTICLE{Aschwanden_2014,
       author = {{Aschwanden}, Markus J. and {Xu}, Yan and {Jing}, Ju},
        title = "{Global Energetics of Solar Flares. I. Magnetic Energies}",
      journal = {\apj},
         year = 2014,
        month = dec,
       volume = {797},
       number = {1},
          eid = {50},
        pages = {50},
          doi = {10.1088/0004-637X/797/1/50},
archivePrefix = {arXiv},
       eprint = {1410.8013},
 primaryClass = {astro-ph.SR},
       adsurl = {https://ui.adsabs.harvard.edu/abs/2014ApJ...797...50A}
}

@ARTICLE{Dahlin_2025,
       author = {{Dahlin}, Joel T. and {Antiochos}, Spiro K. and {DeVore}, C. Richard and {Wyper}, Peter F. and {Qiu}, Jiong},
        title = "{Determining the 3D Dynamics of Solar Flare Magnetic Reconnection}",
      journal = {arXiv e-prints},
         year = 2025,
        month = apr,
          eid = {arXiv:2504.00913},
        pages = {arXiv:2504.00913},
          doi = {10.48550/arXiv.2504.00913},
archivePrefix = {arXiv},
       eprint = {2504.00913},
 primaryClass = {astro-ph.SR},
       adsurl = {https://ui.adsabs.harvard.edu/abs/2025arXiv250400913D}
}

@ARTICLE{Janvier_2015,
       author = {{Janvier}, M. and {Aulanier}, G. and {D{\'e}moulin}, P.},
        title = "{From Coronal Observations to MHD Simulations, the Building Blocks for 3D Models of Solar Flares (Invited Review)}",
      journal = {\solphys},
         year = 2015,
        month = dec,
       volume = {290},
       number = {12},
        pages = {3425-3456},
          doi = {10.1007/s11207-015-0710-3},
archivePrefix = {arXiv},
       eprint = {1505.05299},
 primaryClass = {astro-ph.SR},
       adsurl = {https://ui.adsabs.harvard.edu/abs/2015SoPh..290.3425J}
}

@ARTICLE{Kretzschmar_2013,
       author = {{Kretzschmar}, M. and {Dominique}, M. and {Dammasch}, I.~E.},
        title = "{Sun-as-a-Star Observation of Flares in Lyman {\ensuremath{\alpha}} by the PROBA2/LYRA Radiometer}",
      journal = {\solphys},
         year = 2013,
        month = aug,
       volume = {286},
       number = {1},
        pages = {221-239},
          doi = {10.1007/s11207-012-0175-6},
archivePrefix = {arXiv},
       eprint = {1210.2169},
 primaryClass = {astro-ph.SR},
       adsurl = {https://ui.adsabs.harvard.edu/abs/2013SoPh..286..221K}
}

@ARTICLE{Freeland_1998,
       author = {{Freeland}, S.~L. and {Handy}, B.~N.},
        title = "{Data Analysis with the SolarSoft System}",
      journal = {\solphys},
         year = 1998,
        month = oct,
       volume = {182},
       number = {2},
        pages = {497-500},
          doi = {10.1023/A:1005038224881},
       adsurl = {https://ui.adsabs.harvard.edu/abs/1998SoPh..182..497F}
}

@INPROCEEDINGS{Pietrow_2024,
       author = {{Pietrow}, Alexander G.~M. and {Pastor Yabar}, Adur},
        title = "{Center-to-limb variation of spectral lines and their effect on full-disk observations}",
    booktitle = {Dynamics of Solar and Stellar Convection Zones and Atmospheres},
         year = 2024,
       editor = {{Getling}, Alexander V. and {Kitchatinov}, Leonid L.},
       series = {IAU Symposium},
       volume = {365},
        month = dec,
        pages = {389-393},
          doi = {10.1017/S174392132300501X},
archivePrefix = {arXiv},
       eprint = {2311.06200},
 primaryClass = {astro-ph.SR},
       adsurl = {https://ui.adsabs.harvard.edu/abs/2024IAUS..365..389P}
}

@ARTICLE{Warmuth_2020,
       author = {{Warmuth}, A. and {Mann}, G.},
        title = "{Thermal-nonthermal energy partition in solar flares derived from X-ray, EUV, and bolometric observations. Discussion of recent studies}",
      journal = {\aap},
         year = 2020,
        month = dec,
       volume = {644},
          eid = {A172},
        pages = {A172},
          doi = {10.1051/0004-6361/202039529},
archivePrefix = {arXiv},
       eprint = {2011.04442},
 primaryClass = {astro-ph.SR},
       adsurl = {https://ui.adsabs.harvard.edu/abs/2020A&A...644A.172W}
}

@ARTICLE{Kontar_2017,
       author = {{Kontar}, E.~P. and {Perez}, J.~E. and {Harra}, L.~K. and {Kuznetsov}, A.~A. and {Emslie}, A.~G. and {Jeffrey}, N.~L.~S. and {Bian}, N.~H. and {Dennis}, B.~R.},
        title = "{Turbulent Kinetic Energy in the Energy Balance of a Solar Flare}",
      journal = {Physical Review Letters},
         year = 2017,
        month = apr,
       volume = {118},
       number = {15},
          eid = {155101},
        pages = {155101},
          doi = {10.1103/PhysRevLett.118.155101},
archivePrefix = {arXiv},
       eprint = {1703.02392},
 primaryClass = {astro-ph.SR},
       adsurl = {https://ui.adsabs.harvard.edu/abs/2017PhRvL.118o5101K}
}

@ARTICLE{Kane_1992,
       author = {{Kane}, S.~R. and {McTiernan}, J. and {Loran}, J. and {Fenimore}, E.~E. and {Klebesadel}, R.~W. and {Laros}, J.~G.},
        title = "{Stereoscopic Observations of a Solar Flare Hard X-Ray Source in the High Corona}",
      journal = {\apj},
         year = 1992,
        month = may,
       volume = {390},
        pages = {687},
          doi = {10.1086/171320},
       adsurl = {https://ui.adsabs.harvard.edu/abs/1992ApJ...390..687K}
}

@ARTICLE{Hudson_2001,
       author = {{Hudson}, H.~S. and {Kosugi}, T. and {Nitta}, N.~V. and {Shimojo}, M.},
        title = "{Hard X-Radiation from a Fast Coronal Ejection}",
      journal = {\apjl},
         year = 2001,
        month = nov,
       volume = {561},
       number = {2},
        pages = {L211-L214},
          doi = {10.1086/324760},
       adsurl = {https://ui.adsabs.harvard.edu/abs/2001ApJ...561L.211H}
}

\end{article} 
\end{document}